\PassOptionsToPackage{table}{xcolor}

\documentclass[
	aps,
	prapplied,
	twocolumn,
	reprint,
	superscriptaddress,
	floatfix,
	citeautoscript,
	longbibliography,
	nofootinbib,
]{revtex4-2}

\usepackage[T1]{fontenc}
\usepackage[utf8]{inputenc}

\usepackage{newtxtext}
\usepackage{newtxmath}

\usepackage{chemformula}
\usepackage{siunitx}
\DeclareSIUnit\amu{u}
\DeclareSIUnit\atp{at.\,\%}
\DeclareSIUnit\angstrom{\text {Å}}
\DeclareSIUnit{\ppma}{ppma}
\usepackage{etoolbox} 
\robustify\dots
\usepackage{graphicx}

\usepackage[
	unicode,
	colorlinks = true,
	allcolors = blue,
]{hyperref}

\usepackage{cleveref}
\crefname{appendix}{Appendix}{Appendices}
\AddToHook{cmd/appendix/before}{\crefalias{section}{appendix}}

\usepackage{microtype}

\usepackage{comment}

\usepackage{glossaries}
\glsdisablehyper

\newacronym{1d}{1D}{one-dimensional}
\newacronym{2d}{2D}{two-dimensional}
\newacronym{3d}{3D}{three-dimensional}

\newacronym{ac}{AC}{alternating current}
\newacronym{aes}{AES}{Auger electron spectroscopy}
\newacronym{afm}{AFM}{atomic force microscopy}
\newacronym{alc}{ALC}{avoided level crossing}
\newacronym{api}{API}{application programming interface}
\newacronym{ariel}{ARIEL}{Advanced Rare Isotope Laboratory}
\newacronym{arpes}{ARPES}{angle-resolved photoemission spectroscopy}
\newacronym{atp}{ATP}{adenosine triphosphate}

\newacronym[sort={b-NMR}]{bnmr}{\ensuremath{\beta}-NMR}{\ensuremath{\beta}-detected nuclear magnetic resonance}
\newacronym[sort={b-NMR2}]{bnmr2}{\ensuremath{\beta}-NMR}{\ensuremath{\beta}-radiation-detected NMR}
\newacronym[sort={b-NQR}]{bnqr}{\ensuremath{\beta}-NQR}{\ensuremath{\beta}-detected nuclear quadrupole resonance}
\newacronym{bca}{BCA}{binary collision approximation}
\newacronym{bcc}{BCC}{body-centred cubic}
\newacronym{bcp}{BCP}{buffered chemical polishing}
\newacronym{bcs}{BCS}{Bardeen-Cooper-Schrieffer}
\newacronym{bpp}{BPP}{Bloembergen-Purcell-Pound}
\newacronym{bsc}{BSC}{\ch{Bi2Se3:Ca}}
\newacronym{btcs}{BTCS}{backward-time, centred-space}
\newacronym{btm}{BTM}{\ch{Bi2Te3:Mn}}
\newacronym{bts}{BTS}{\ch{Bi2Te2Se}}

\newacronym{camp}{CAMP}{control and monitor program}
\newacronym{ccd}{CCD}{charge-coupled device}
\newacronym{cdw}{CDW}{charge density wave}
\newacronym{cgs}{CGS}{centimetre-gram-second system of units}
\newacronym{cmms}{CMMS}{Centre for Molecular and Materials Science}
\newacronym{cn}{CN}{Crank-Nicolson}
\newacronym{codata}{CODATA}{Committee on Data for Science and Technology}
\newacronym{cpu}{CPU}{central processing unit}
\newacronym{create}{CREATE}{Collaborative Research and Training Experience Program}
\newacronym{ctcs}{CTCS}{centred-time, centred-space}
\newacronym{cw}{CW}{continuous wave}

\newacronym{daq}{DAQ}{data acquisition}
\newacronym{dc}{DC}{direct current}
\newacronym{dft}{DFT}{density functional theory}
\newacronym{dos}{DOS}{density of states}
\newacronym{dqt}{DQT}{double-quantum transition}

\newacronym{efg}{EFG}{electric field gradient}
\newacronym{emim-ac}{EMIM-Ac}{1-ethyl-3-methylimidazolium acetate}
\newacronym{emim-dca}{EMIM-DCA}{1-ethyl-3-methylimidazolium dicyanamide}
\newacronym{ep}{EP}{electro-polishing}
\newacronym{epr}{EPR}{electron paramagnetic resonance}
\newacronym{esr}{EPR}{electron spin resonance}
\newacronym{endor}{ENDOR}{electron nuclear double resonance}
\newacronym{epics}{EPICS}{Experimental Physics and Industrial Control System}

\newacronym{fcc}{FCC}{face-centred cubic}
\newacronym{fft}{FFT}{fast Fourier transform}
\newacronym{fom}{FoM}{figure of merit}
\newacronym{ftcs}{FTCS}{forward-time, centred-space}
\newacronym{fwhm}{FWHM}{full width at half maximum}

\newacronym{gga}{GGA}{generalized gradient approximation}
\newacronym{gl}{GL}{Ginzburg-Landau}

\newacronym{hb}{HB}{hole-burning}
\newacronym{hfqs}{HFQS}{high-field \ensuremath{Q} slope}
\newacronym{hv}{HV}{high-voltage}
\newacronym{hwhm}{HWHM}{half width at half maximum}

\newacronym{iaea}{IAEA}{International Atomic Energy Agency}
\newacronym{icru}{ICRU}{International Commission on Radiation Units and Measurements}
\newacronym{il}{IL}{ionic liquid}
\newacronym{is}{IS}{impedance spectroscopy}
\newacronym{isac}{ISAC}{isotope separator and accelerator}
\newacronym{isol}{ISOL}{isotope separation online}
\newacronym{isosim}{IsoSiM}{Isotopes for Science and Medicine}

\newacronym{lcao}{LCAO}{linear combination of atomic orbitals}
\newacronym{lda}{LDA}{local density approximation}
\newacronym{led}{LED}{light-emitting diode}
\newacronym{leis}{LEIS}{low-energy ion scattering}
\newacronym{lib}{LIB}{lithium-ion battery}
\newacronym{lsat}{LSAT}{\ch{(La,Sr)(Al,Ta)O3}}

\newacronym{mas}{MAS}{magic angle spinning}
\newacronym{mpms}{MPMS}{magnetic property measurement system}
\newacronym{mbe}{MBE}{molecular beam epitaxy}
\newacronym{md}{MD}{molecular dynamics}
\newacronym{midas}{MIDAS}{Maximum Integrated Data Acquisition System}
\newacronym{mit}{MIT}{metal-insulator transition}
\newacronym{mnr}{MNR}{Meyer-Neldel rule}
\newacronym{mqt}{mqt}{multi-quantum transition}
\newacronym{mud}{MUD}{muon data}
\newacronym{ms}{MS}{mass spectrometry}

\newacronym{nbm}{NBM}{neutral beam monitor}
\newacronym{neb}{NEB}{nudged elastic band}
\newacronym{nim}{NIM}{nuclear instrumentation module}
\newacronym{nlme}{NLME}{non-linear Meissner effect}
\newacronym{nmr}{NMR}{nuclear magnetic resonance}
\newacronym{no}{NO}{nuclear orientation}
\newacronym{nqr}{NQR}{nuclear quadrupole resonance}
\newacronym{nrc}{NRC}{National Research Council of Canada}
\newacronym{nserc}{NSERC}{Natural Sciences and Engineering Research Council of Canada}

\newacronym{oa}{OA}{optical absorption}

\newacronym{pac}{PAC}{perturbed angular correlation}
\newacronym{pad}{PAD}{perturbed angular distribution}
\newacronym{pas}{PAS}{principle axis system}
\newacronym{pchip}{PCHIP}{piecewise cubic Hermite interpolating polynomial}
\newacronym{pdf}{PDF}{probability density function}
\newacronym{pld}{PLD}{pulsed laser deposition}
\newacronym{ppms}{PPMS}{physical property measurement system}
\newacronym{psi}{PSI}{Paul Scherrer Institute}

\newacronym{qens}{QENS}{quasielastic neutron scattering}
\newacronym{ql}{QL}{quintuple layer}
\newacronym{qo}{QO}{quantum oscillations}

\newacronym{rbs}{RBS}{Rutherford backscattering}
\newacronym{rf}{RF}{radio frequency}
\newacronym{rheed}{RHEED}{reflection high-energy electron diffraction}
\newacronym{rib}{RIB}{radioactive ion beam}
\newacronym{rkky}{RKKY}{Ruderman–Kittel–Kasuya–Yosida}
\newacronym{rrr}{RRR}{residual-resistivity ratio}
\newacronym{rtil}{RTIL}{room temperature ionic liquid}

\newacronym{sae}{SAE}{spin-alignment echo}
\newacronym{sans}{SANS}{small angle neutron scattering}
\newacronym{si}{SI}{International System of Units}
\newacronym{sims}{SIMS}{secondary ion mass spectrometry}
\newacronym{slr}{SLR}{spin-lattice relaxation}
\newacronym{sms}{S\ensuremath{\mu}S}{Swiss Muon Source}
\newacronym[sort={S/N}]{snr}{\textit{S}/\textit{N}}{signal-to-noise ratio}
\newacronym{squid}{SQUID}{superconducting quantum interference device}
\newacronym{srf}{SRF}{superconducting radio frequency}
\newacronym{srim}{SRIM}{Stopping and Range of Ions in Matter}
\newacronym{ss}{SS}{superconductor-superconductor}
\newacronym{ssid}{SSID}{solid-state ionic device}
\newacronym{ssr}{SSR}{spin-spin relaxation}
\newacronym{stm}{STM}{scanning tunnelling microscopy}
\newacronym{sts}{STS}{scanning tunnelling spectroscopy}

\newacronym{ti}{TI}{topological insulator}
\newacronym{trim}{TRIM}{Transport and Range of Ions in Matter}
\newacronym{tss}{TSS}{topological surface state}
\newacronym{tmd}{TMD}{transition metal dichalcogenide}

\newacronym{uhv}{UHV}{ultra-high vacuum}

\newacronym{vdw}{vdW}{van der Waals}
\newacronym{vft}{VFT}{Vogel-Fulcher-Tammann}

\newacronym{xps}{XPS}{x-ray photoelectron spectroscopy}
\newacronym{xrd}{XRD}{x-ray diffraction}
\newacronym{xrr}{XRR}{x-ray reflection}

\newacronym{ybco}{YBCO}{\ch{YBa2Cu3O_{6+x}}}
\newacronym{ysz}{YSZ}{yttria-stabilized zirconia}

\newacronym[sort={muSR}]{musr}{\ensuremath{\mu}SR}{muon spin spectroscopy}
\newacronym{alc-musr}{ALC-\ensuremath{\mu}SR}{avoided level crossing muon spin rotation}
\newacronym{le-musr}{LE-\ensuremath{\mu}SR}{low-energy muon spin spectroscopy}

\newacronym{lf-musr}{LF-\ensuremath{\mu}SR}{longitudinal field muon spin rotation}
\newacronym{rf-musr}{RF-\ensuremath{\mu}SR}{radio frequency muon spin rotation}
\newacronym{tf-musr}{TF-\ensuremath{\mu}SR}{transverse field muon spin rotation}
\newacronym{zf-musr}{ZF-\ensuremath{\mu}SR}{zero field muon spin rotation}

\begin{document}

\title{
	Insight into SRF cavity performance from simulations
	of Nb's surface oxide dissolution and diffusion 
}

\author{Ryan~M.~L.~McFadden}
\email[E-mail: ]{rmlm@triumf.ca}
\affiliation{TRIUMF, 4004 Wesbrook Mall, Vancouver, BC V6T~2A3, Canada}
\affiliation{Department of Physics and Astronomy, University of Victoria, 3800 Finnerty Road, Victoria, BC V8P~5C2, Canada}

\author{Rowan~Becker}
\affiliation{Department of Physics and Astronomy, University of Victoria, 3800 Finnerty Road, Victoria, BC V8P~5C2, Canada}

\author{Tobias~Junginger}
\email[E-mail: ]{junginger@uvic.ca}
\affiliation{TRIUMF, 4004 Wesbrook Mall, Vancouver, BC V6T~2A3, Canada}
\affiliation{Department of Physics and Astronomy, University of Victoria, 3800 Finnerty Road, Victoria, BC V8P~5C2, Canada}

\date{\today}

\begin{abstract}
We report simulations of the dissolution and diffusion of Nb’s surface oxide layer in vacuum. While this
chemical doping process is important for the surface preparation of Nb superconducting radio frequency (SRF)
cavities --- common components of particle accelerators --- quantitatively linking the resulting oxygen distributions
to superconducting performance remains challenging.
In this work,
we simulate the reaction-diffusion process numerically
for treatment temperatures $T = \qtyrange{50}{200}{\celsius}$
and times
$t = \qtyrange{0.5}{120}{\hour}$,
and calculate the effect of the spatially inhomogeneous oxygen doping
on \ch{Nb}'s superconducting properties. We find that oxygen doping redistributes the Meissner screening current, reducing its value at the surface and shifting its maximum several nanometres into the material. These results provide a microscopic link between oxygen diffusion profiles and the electromagnetic response of Nb relevant for SRF cavity operation. This work provides a quantitative framework linking oxygen diffusion profiles to superconducting performance and establishes a foundation for future studies involving time-dependent and multi-step heat treatment protocols.
\end{abstract}

\maketitle
\glsresetall

\section{
	Introduction
	\label{sec:introduction}
}

The operation of \gls{srf} cavities fabricated using \ch{Nb}
---
common components of particle accelerators
---
relies on the elemental superconductor maintaining its Meissner state under
\gls{rf} electromagnetic fields,
which are used to convert stored electromagnetic energy
to
kinetic energy for the accelerated (charged) particles~\cite{2023-Padamsee-SRTA}.
In the interest of maximizing \gls{srf} cavity utility
(i.e., largest accelerating gradients, smallest accelerator structures, etc.),
utilizing electromagnetic fields up to \ch{Nb}'s fundamental
material limits
(i.e., its superheating field $B_{\mathrm{sh}} \approx \qty{240}{\milli\tesla}$) 
is crucial.
While achieving this in practice remains challenging
(see, e.g.,~\cite{2025-Kubo-JJAP-64-018002}),
most advancements toward this goal have been achieved through empirical heat treatment ``recipes,''
such as ``low-$T$''~\cite{2004-Ciovati-JAP-96-1591}
or ``mid-$T$''\cite{2020-Posen-PRA-13-014024,2021-Ito-PTEP-2021-071G01}
baking in vacuum.

Among their attributes,
these heat treatments cause the dissolution
and
diffusion~\cite{1990-King-TSF-192-351,2006-Ciovati-APL-89-022507,2021-Lechner-APL-119-082601,2024-Lechner-JAP-135-133902}
of the metal's
native surface oxide layer
(i.e., \ch{NbO_{x}}~\cite{1987-Halbritter-APA-43-1}),
leading to modest,
chemical doping of the near-surface region by interstitial oxygen.
Depending on the treatment ``recipe,''
the doping may be spatially inhomogeneous
and
confined to depths on the order of 10s or 100s of nanometers.
While quantifying the level of doping is possible
using,
for example,
\gls{sims}~\cite{2026-McFadden-PRB-113-L060508},
measuring its impact on \ch{Nb}'s electromagnetic response
\emph{on the same length scale}
remain challenging.
Though some progress has been made recently using \gls{le-musr}
(see, e.g.,~\cite{2023-McFadden-PRA-19-044018,2024-McFadden-APL-124-086101,2024-McFadden-AIPA-14-095320}),
having additional vantage points 
would be beneficial for the \gls{srf} community.

Simulating the doping process and its effect on \ch{Nb}'s superconductivity
is an attractive avenue for improving our understanding of the
\emph{microscopic} details important for function in \emph{macroscopic} cavity resonators.
On practical grounds, such an approach is appealing:
fabricating and testing full cavity structures is both onerous and costly,
whereas continuum simulations of the doping process are inexpensive
and
expedient
(i.e., amenable to running on a personal computer).
At a fundamental level,
the main strength of simulations is their potential to help bridge the gap between the micro and macro world.
Unfortunately,
there is great ambiguity in the details bridging these scales
owing to the multitude of factors affecting cavity performance~\cite{2023-Padamsee-SRTA}.
Nonetheless,
progress continues to be made on this front~\cite{2024-Lechner-JAP-135-133902,2025-Kubo-JJAP-64-018002},
connecting,
for example,
the oxygen diffusion length to cavity resonator observables~\cite{2025-Bate-SST-38-025003}.
Though some simulation work on dissolution and diffusion has been done
(see, e.g.,~\cite{2006-Ciovati-APL-89-022507,2021-Lechner-APL-119-082601,2024-Lechner-JAP-135-133902}),
it is often used only as a minor accompanying detail to complement experimental work
(see, e.g.,~\cite{2024-Prudnikava-SST-37-075007,2025-Tamashevich-SST-38-045006}).
Moreover,
simulations relating the microscopic doping levels to electromagnetic properties,
especially \ch{Nb}'s supercurrent density,
remains relatively unexplored~\cite{2021-Lechner-APL-119-082601,2020-Checchin-APL-117-032601,2024-Lechner-JAP-135-133902}.

To gain better insight into the effect of the treatment process on
supercurrent densities near \ch{Nb}'s surface,
in this work we simulate the reaction-diffusion system
numerically under conditions commonly used in \gls{srf} cavity fabrication.
Using the best available experimental data as inputs for our model of the reaction-diffusion 
system,
we simulate depth-dependent oxygen dopant profiles for fixed vacuum annealing temperatures and times
relevant for \ch{Nb} \gls{srf} cavities.
From the profiles,
we calculate their effect on \ch{Nb}'s superconducting properties. From the simulated oxygen concentration profiles, we calculate the resulting spatial dependence of the electron mean free path, magnetic penetration depth, critical current density, and Meissner screening currents. We find that oxygen diffusion substantially modifies the near-surface current distribution, reducing the surface screening current and shifting its maximum several nanometres beneath the surface. By systematically exploring treatment temperatures and durations, we identify regions of parameter space that may offer improved performance relative to standard SRF cavity baking recipes.

The remainder of this manuscript is organized as follows.
In \Cref{sec:simulations},
a detailed description of our simulations is provided,
along with how \ch{Nb}'s superconducting properties are calculated from their output.
A breakdown of our results is given in \Cref{sec:results},
followed by a discussion of their significance in \Cref{sec:discussion}. 
Conclusions drawn from our work are summarized in \Cref{sec:conclusions}.
Additional technical information and supplementary simulation results are also provided in
\Cref{sec:oxygen-diffusion,sec:crank-nicolson,sec:rate-constants,sec:additional-simulations}.

\section{
	Simulations
	\label{sec:simulations}
}

To model the dissolution and diffusion process,
we treat the phenomenon as a \gls{1d} reaction-diffusion system.
This approach has been shown to be successful for quantifying the degree
of oxygen doping in \ch{Nb} following vacuum annealing
(see, e.g.,~\cite{2006-Ciovati-APL-89-022507,2024-Lechner-JAP-135-133902}),
and
we impliment a version that works for \emph{arbitrary} simulation conditions
through the use of fast, stable numeric techniques.
In the remainder of this section,
we outline the kinetics of surface oxide dissolution in \Cref{sec:simulations:dissolution},
followed by the diffusion on interstitial oxygen in \Cref{sec:simulations:diffusion}.
Material properties derived from the resulting oxygen interstices
are detailed in \Cref{sec:simulations:properties},
with an overview of the full approach given in \Cref{sec:simulations:approach}.

\subsection{
	Surface Oxide Dissolution
	\label{sec:simulations:dissolution}
}

The dissolution of oxygen from \ch{Nb}'s native surface oxide layer
(see, e.g.,~\cite{1987-Halbritter-APA-43-1}) can be (qualitatively) understood
as the thermally mediated reduction of \ch{Nb}'s nominal valence,
going from 5+ (in the pentoxide) to 0 (in the metal),
as determined by \gls{xps}~\cite{1990-King-TSF-192-351}.
In vacuum,
this happens in an irreversible step-wise manner according to:
\begin{align}
	\label{eq:reaction-pentoxide}
	\ch{Nb2O5} & \xrightarrow{k_{1}} 2\ch{NbO2} + \ch{O} , \\
	\label{eq:reaction-dioxide}
	\ch{NbO2}  & \xrightarrow{k_{2}} \ch{NbO} + \ch{O} , \\
	\label{eq:reaction-monoxide}
	\ch{NbO}   & \xrightarrow{k_{3}} \ch{Nb} + \ch{O} ,
\end{align}
where the $k_{i}$s denote rate constants for the kinetic process.
In practice,
$k_{3} \approx \qty{0}{\per\second}$
(up to at least \qty{1000}{\kelvin})~\cite{1990-King-TSF-192-351},
greatly simplifying the (first-order) rate equations for the above reactions:
\begin{align}
	\label{eq:diff-Nb2O5}
	\frac{ \mathrm{d} }{\mathrm{d}t} [\ch{Nb2O5}](t) & = -k_{1} [\ch{Nb2O5}] , \\
	\label{eq:diff-NbO2}
	\frac{ \mathrm{d} }{\mathrm{d}t} [\ch{NbO2}](t) & = -k_{2} [\ch{NbO2}] + 2 k_{1} [\ch{Nb2O5}] , \\
	\label{eq:diff-NbO}
	\frac{ \mathrm{d} }{\mathrm{d}t} [\ch{NbO}](t) & \approx k_{2} [\ch{NbO2}] , \\
	\label{eq:diff-O}
	\frac{ \mathrm{d} }{\mathrm{d}t} [\ch{O}](t) & \approx k_{1} [\ch{Nb2O5}] + k_{2} [\ch{NbO2}] ,
\end{align}
where $[\cdots]$s are used to denote the concentration of each chemical species
and
$t$ denotes time.
These equations have the general solutions:
\begin{widetext}
\begin{align}
	\label{eq:Nb2O5-concentration}
	[\ch{Nb2O5}](t) & = [\ch{Nb2O5}]_{0} \exp ( -k_{1} t ) , \\
	\label{eq:NbO2-concentration}
	[\ch{NbO2}](t) & = \frac{ \mathcal{C}(t) + 2 [\ch{Nb2O5}]_{0} k_{1} \exp (-k_{1}t)  }{ k_{2} - k_{1} } , \\
	\label{eq:NbO-concentration}
	[\ch{NbO}](t) & = [\ch{NbO}]_{\mathrm{max}} - \frac{ \mathcal{C}(t) + 2 [\ch{Nb2O5}]_{0} k_{2} \exp (-k_{1}t)  }{ k_{2} - k_{1} } , \\
	\label{eq:O-concentration}
	[\ch{O}](t) & = [\ch{O}]_{\mathrm{max}} - \frac{ \mathcal{C}(t) + [\ch{Nb2O5}]_{0}(3k_{2} - k_{1}) \exp(-k_{1}t) }{ k_{2} - k_{1} },
\end{align}
\end{widetext}
where
$[\cdots]_{0}$ denotes the initial concentration of each chemical species at
$t = \qty{0}{\second}$,
\begin{widetext}
\begin{equation}
	\label{eq:common-concentration}
	\mathcal{C}(t) \equiv \left \{ [\ch{NbO2}]_{0} k_{2} - k_{1} \left ([\ch{NbO2}]_{0} + 2 [\ch{Nb2O5}]_{0} \right ) \right \} \exp (-k_{2}t) 
\end{equation}
\end{widetext}
denotes the ``common'' term in
\Cref{eq:NbO2-concentration,eq:NbO-concentration,eq:O-concentration},
and
\begin{align}
	\label{eq:NbO-max-concentration}
	[\ch{NbO}]_{\mathrm{max}} & \equiv 2[\ch{Nb2O5}]_{0} + [\ch{NbO2}]_{0} + [\ch{NbO}]_{0} , \\
	\label{eq:O-max-concentration}
	[\ch{O}]_{\mathrm{max}} & \equiv 3[\ch{Nb2O5}]_{0} + [\ch{NbO2}]_{0} + [\ch{O}]_{0} ,
\end{align}
define the maximum concentrations obtainable for
\ch{NbO} and \ch{O},
respectively
(i.e., $\lim_{t \rightarrow \infty} [ \ch{X} ](t) = [\ch{X}]_{\mathrm{max}}$,
where $\ch{X} \equiv \ch{NbO}, \ch{O}$).
These expressions are similar to
those listed in Ref.~\cite{2024-Lechner-JAP-135-133902},
and
consistent with Ref.~\cite{2006-Ciovati-APL-89-022507}
in the limit that $k_{2}$ is negligible.

Typical of reaction rate constants,
each of the $k_{i}$s in
\Cref{eq:Nb2O5-concentration,eq:NbO2-concentration,eq:NbO-concentration,eq:O-concentration,eq:common-concentration}
follow an Arrhenius temperature dependence~\cite{1990-King-TSF-192-351,2024-Prudnikava-SST-37-075007,2021-Lechner-APL-119-082601,2024-Lechner-JAP-135-133902}:
\begin{equation}
	\label{eq:kinetic-arrhenius}
	k_{i} = A_{0,i} \exp \left ( -\frac{ E_{A,i} }{ R T } \right ),
\end{equation}
where
$A_{0,i}$ is a preexponential factor related the reaction's ``attempt'' frequency,
$E_{A,i}$ is the (classical) energy barrier that must be overcome along the reaction
``pathway,''
$T$ is the absolute temperature,
and
$R = \qty{8.314 462 618 \dots}{\joule\per\mol\per\kelvin}$ is the molar gas constant~\cite{2021-Tiesinga-RMP-93-025010}.
Values for these quantities have been determined,
for example,
using \gls{xps}~\cite{1990-King-TSF-192-351,2024-Prudnikava-SST-37-075007}
and
\gls{sims}~\cite{2021-Lechner-APL-119-082601,2024-Lechner-JAP-135-133902}.
In this work,
we take the \emph{average} of these literature values as the best estimate
for each $k_{i}$'s $T$-dependence.
Specifically,
we use:
$A_{0,1} = \qty{9.75e8}{\per\second}$ and $E_{A,1} = \qty{133.7}{\kilo\joule\per\mol}$ for $k_{1}$;
and
$A_{0,2} = \qty{1.5e11}{\per\second}$ and $E_{A,2} = \qty{177}{\kilo\joule\per\mol}$ for $k_{2}$.
A full discussion on this choice is given in \Cref{sec:rate-constants}

Important for our model is the rate of interstitial oxygen production,
which acts as the source term in the reaction-diffusion system
(see \Cref{sec:simulations:diffusion}).
Differentiating \Cref{eq:O-max-concentration} with respect to time,
we obtain:
\begin{widetext}
\begin{equation}
	\label{eq:O-production-rate}
	\frac{\mathrm{d}}{\mathrm{d}t} [\ch{O}](t) = \frac{ k_{2} \mathcal{C}(t) + k_{1} [\ch{Nb2O5}]_{0} \left ( 3 k_{2} - k_{1} \right ) \exp \left ( -k_{1} t \right ) }{ k_{2} - k_{1} } ,
\end{equation}
\end{widetext}
in agreement with with Ref.~\cite{2024-Lechner-JAP-135-133902}.
Again,
in the limit that $k_{2} \rightarrow \qty{0}{\per\second}$,
\Cref{eq:O-production-rate} reduces to the expression given in Ref.~\cite{2006-Ciovati-APL-89-022507}.

\subsection{
	Interstitial Oxygen Diffusion
	\label{sec:simulations:diffusion}
}

The migration of interstitial oxygen, initially localized near \ch{Nb}'s surface,
to deeper within the metal's interior is mediated by (classical) diffusion of the atom.
Taking into account the time-dependent nature in which oxygen is liberated from surface oxide
(see \Cref{sec:simulations:dissolution}),
this process can be modeled using the reaction-diffusion equation~\cite{2024-Lechner-JAP-135-133902,2006-Ciovati-APL-89-022507}:
\begin{equation}
        \label{eq:reaction-diffusion}
        \frac{\partial }{\partial t} [\ch{O}](x,t) = D \frac{\partial^2}{\partial x^2} [\ch{O}](x,t) + q(x, t)
\end{equation}
where $[\ch{O}](x,t)$ is the oxygen concentration as a function of depth $x$ and time $t$,
and
\begin{equation}
   \label{eq:source}
   q(x, t) \equiv \frac{\mathrm{d}}{\mathrm{d}t}[\ch{O}](t) \cdot \delta (x) ,
\end{equation}
is the ``source'' term
accounting for the rate in which oxygen is introduced to the system [\Cref{eq:O-production-rate}],
assumed to be localized at the surface (i.e., as a plane source).
Note that this latter approximation in \Cref{eq:source} is
reasonable~\cite{2006-Ciovati-APL-89-022507,2021-Lechner-APL-119-082601,2024-Lechner-JAP-135-133902,2025-Tamashevich-SST-38-045006},
given the limited extent of \ch{Nb}'s native surface oxide
(thickness $\lesssim \qty{ 5}{\nano\meter}$ --- see, e.g.,~\cite{1987-Halbritter-APA-43-1}).
The remaining term $D$ in \Cref{eq:reaction-diffusion}
is the diffusion coefficient or diffusivity,
which accounts for the haste in which the diffusing species ``spreads'' from
concentrated regions.

Analogous to the $T$-dependence of the $k_{i}$s in \Cref{sec:simulations:dissolution},
$D$ follows a similar
thermally activated relation:
\begin{equation}
    \label{eq:diffusion-arrhenius}
    D = D_{0} \exp \left ( - \frac{E_{A, D}}{ R T } \right ) ,
\end{equation}
where $D_{0}$ is a prefactor,
$E_{A,D}$ is the (classic) migration barrier,
and
the remaining terms take the same meaning as in \Cref{eq:kinetic-arrhenius}.
The diffusion of interstitial oxygen in \ch{Nb} has been extensively studied,
with most of the measurements summarized in Ref.~\cite{1990-LeClaire-LBIII-26-471}.
Following the analysis in Refs.~\cite{1977-Boratto-SM-11-709,1980-Boratto-MSE-43-97},
we adopt values of 
$D_{0} = \qty{0.59e-6}{\meter\squared\per\second}$
and
$E_{A,D} = \qty{109.7}{\kilo\joule\per\mole}$
for our simulations.
Note that this choice implicitly assumes at the microscopic level
oxygen migration takes place between (quasi)octahedral interstitial sites in \ch{Nb}'s \gls{bcc}
lattice (see, e.g.,~\cite{1965-Beshers-JAP-36-290}). 
Further discussion of this is given in
\Cref{sec:oxygen-diffusion}.

To find the depth-dependent oxygen doping profile resulting from the dissolution
and diffusion process,
it is necessary to solve the reaction-diffusion equation.
Although this may be done analytically in select cases~\cite{1975-Crank-TMOD-2},
finding solutions to the partial differential equation \emph{numerically}
is a viable general alternative
(see, e.g., Ref.~\cite{2003-Hundsdorfer-SSCM-33}).
In particular,
the \gls{cn} (semi-implicit) method~\cite{1947-Crank-MPCPS-43-50,1975-Crank-TMOD-2}
is one such approach that offers both simplicity and numerical stability,
and we briefly sketch it below.

The \gls{cn} method 
discretizes the (continuous) diffusion problem onto an $N_t \times N_x$ grid of spatial
and temporal points with indices $n_i = 0, 1, 2, \dots , N_{i} - 2, N_{i} - 1$,
wherein all derivatives are calculated numerically
(appropriately accounting from the system's boundary conditions).
In doing so,
it is possible to recast the problem to a set of linear equations,
which may be succinctly expressed using vectors and matrices.
Starting from a known (i.e., defined) condition at time index $n_{t}$,
(discrete) spatial solutions for $[\ch{O}](x)$ can be generated at
$n_{t}+1$ by an expression of the form:
\begin{equation}
    \label{eq:crank-nicolson-solution}
    \mathbf{O}^{n_{t} + 1} = \mathsf{A}^{-1} \left ( \mathsf{B} \mathbf{O}^{n_{t}} + \Delta t \mathbf{Q}^{n_{t}} \right ) ,
\end{equation}
where $\mathbf{O}$ and $\mathbf{Q}$ are vectors representing
discretized forms of $[\ch{O}](x)$ and $q(x)$,
$\mathsf{A}$ and $\mathsf{B}$ are (sparse) matrices that account for the diffusion
between time points,
and
$\Delta t$ is the time-step between indices $n_{t}$ and $n_{t}+1$.
Thus,
\Cref{eq:crank-nicolson-solution} 
may be applied iteratively to ``evolve'' the spatial form of the oxygen
dopant profile for arbitrary $t$.
A more complete description of the \gls{cn} method~\cite{1947-Crank-MPCPS-43-50,1975-Crank-TMOD-2}
is provided in \Cref{sec:crank-nicolson}.

\subsection{
	Derived Material Properties
	\label{sec:simulations:properties}
}

The defects produced from the reaction-diffusion process
(i.e., interstitial oxygen atoms)
modify the metal's electronic properties
over the profile's spatial extent,
which affect features of its superconducting state.
Chiefly, the Meissner response is modified resulting in a reduction in the supercurrent density.
A key detail in the dissolution/diffusion process is that,
for typical baking ``recipes,''
the resulting defect profiles are spatially \emph{inhomogeneous}
which,
as a corollary,
should impart a depth-dependent character to all derived quantities.
In the following,
we explore these connections explicitly.

\subsubsection{
	Electron Mean-Free-Path
	\label{sec:simulations:properties:mean-free-path}
}

The electron mean-free-path 
$\ell$
defines the average distance the charge carrier
travels in a material before its energy/direction is (substantially) altered by a ``scattering'' event
(e.g., ``colliding'' with an impurity atom or phonon).
In this sense,
$\ell$ is directly related to the abundance of lattice defects
(e.g., impurity atoms, vacancies, dislocations, grain boundaries, etc.)
present in \ch{Nb} metal.
In the limit of $T \rightarrow \qty{0}{\kelvin}$,
the most important contribution is generally impurity atoms~\cite{1981-Schulze-JM-33-33,2000-Koethe-MTJIM-41-7},
which act as electronic scattering centres
and
raise \ch{Nb}'s residual resistivity.
This behavior can be expressed through the empirical relationship between
\ch{Nb}'s residual resistivity $\rho_{e}$ and $\ell$~\cite{1968-Goodman-JPF-29-240}:
\begin{equation}
	\label{eq:mfp-rrr}
	\ell(x) = \frac{\sigma_{e}}{\rho_{e}(x)} ,
\end{equation}
where $\sigma_{e} = \qty{3.7e-16}{\ohm\meter\squared}$~\cite{1968-Goodman-JPF-29-240} is an empirical material constant relating the electron mean free path to the residual resistivity of Nb~\cite{1968-Goodman-JPF-29-240}.
Noting that $\rho_{e}(x)$ is the sum of contributions from all
impurity atoms,
it may be expressed generally as:
\begin{equation}
	\label{eq:residual-resistivity}
	\rho_{e}(x) = \sum_{i} a_{i} \cdot [i](x) ,
\end{equation}
where $i$ denotes the impurity species
and $a_{i}$ denoting the (linear) proportionality with
concentration $[i](x)$.
For common impurity atoms in \ch{Nb}
(e.g., \ch{C}, \ch{N}, \ch{O}, etc.),
the empirical proportionality constants are known,
with
$a_{\ch{O}} = \qty{4.5 \pm 0.3 e-12}{\ohm\meter\per\ppma}$~\cite{1976-Schulze-ZM-67-737}.
Thus,
from knowledge of the oxygen concentration profile,
one may calculate the spatial dependence of $\ell$ using
\Cref{eq:mfp-rrr,eq:residual-resistivity}.

As $\ell(x)$ influences some of the salient properties of a superconductor
(e.g., its Meissner response),
knowledge of its value is of particular interest for \gls{srf} \ch{Nb}.
We consider some of these ``connected'' properties below.

\subsubsection{
	Magnetic Penetration Depth
	\label{sec:simulations:properties:magnetic-penetration-depth}
}

In brief,
for a superconductor in its Meissner state,
the magnetic penetration depth $\lambda$ 
describes the characteristic length scale
(typically 10s to 100s of nanometers)
over which external magnetic flux may ``creep'' below the surface
before being completely screened deeper in its interior.
In the limit that the superconductor's electrodynamics
are purely \emph{local},
the London model~\cite{1935-London-PRSLA-149-71}
adequately describes the magnetic-flux-screening
and
$\lambda$ is simply the exponential decay
length.\footnote{In \emph{nonlocal} superconductors (i.e., those with coherence lengths $\xi_{0} \gg \lambda$), more complicated expressions are required to describe the Meissner effect and the ``effective'' screening length (see, e.g., Refs.~\cite{1953-Pippard-PRSLA-216-547,1957-Bardeen-PR-108-1175,1971-Halbritter-ZP-243-201,1996-Tinkham-IS-2}); however, as nonlocal effects are small even in ``clean'' \ch{Nb} (i.e., where differences between $\xi_{0}$ and $\lambda$ are greatest)~\cite{2026-McFadden-PRB-113-L060508}, the local-limit expressions used in this work provide an excellent description of the metal's electrodynamics.}
In this context,
its temperature-dependence
can be accurately approximated using~\cite{1996-Tinkham-IS-2,2005-Suter-PRB-72-024506}:
\begin{equation}
	\label{eq:lambda-two-fluid}
	\lambda(T, x) \approx \begin{cases}
		\dfrac{ \lambda_{0}(x) }{ \sqrt{1 - \left ( T / T_{c} \right )^{4} } }, & T \leq T_{c} , \\
		+\infty, & T > T_{c} ,
	\end{cases}
\end{equation}
where 
$T$ is the absolute temperature,
$T_{c} \approx \qty{9.25}{\kelvin}$ is \ch{Nb}'s critical temperature~\cite{1966-Finnemore-PR-149-231,2022-Turner-SR-12-5522},
and
$\lambda_{0}(x)$ is the penetration depth's value extrapolated to \qty{0}{\kelvin}.
Where $\lambda(T,x)$ becomes sensitive to the presence of impurities is through the
term $\lambda_{0}(x)$,
which scales according to the inverse square root of $\ell$
(i.e., lower impurity concentrations yield \emph{longer} $\ell$s,
which correspond to \emph{shorter} $\lambda$s).
More explicitly~\cite{1959-Miller-PR-113-1209,1971-Halbritter-ZP-243-201},
\begin{equation}
	\label{eq:lambda-impurity}
	\lambda_{0}(x) \approx \lambda_{L} \sqrt{ 1 + \frac{ \pi \xi_{0} }{ 2 \ell(x) } } ,
\end{equation}
where $\lambda_{L}$ is the London penetration depth
(i.e., a pure material's intrinsic magnetic screening length)
and
$\xi_{0}$ is the Pippard~\cite{1953-Pippard-PRSLA-216-547}/\gls{bcs}~\cite{1957-Bardeen-PR-108-1175}
coherence length
(i.e., the spatial extent of Cooper pairs in the superconducting state).\footnote{Note that the factor $\pi/2$ included in \Cref{eq:lambda-impurity} is in accord with \gls{bcs} theory~\cite{1957-Bardeen-PR-108-1175} for impure superconductors~\cite{1959-Miller-PR-113-1209,1971-Halbritter-ZP-243-201}. Other versions of this expression exist is the literature without this factor, where the authors have (implicitly) assumed a different definition of $\xi_{0}$.}

\subsubsection{
	Non-linear Meissner Effect
	\label{sec:simulations:properties:nlme}
}

Empirically,
it is well-known that $\lambda(T, x)$ is affected by an applied magnetic field $B_{0}$,
a result that was first understood theoretically using
\gls{gl} theory~\cite{1950-Ginzburg-ZETF-20-1064,1965-Ginzburg-CPLDL-73-546},
with later developments extending it to unconventional superconductors
(see, e.g.,~\cite{1995-Xu-PRB-51-16233})
Qualitatively,
the effect accounts for $\lambda(T, x)$'s quadratic increase with increasing $B_{0}$.
Generally, the effect is small for low $B_{0}$,
but it can become significant when $B_{0}$ is
comparable to the thermodynamic critical field.

Quantitatively,
the \gls{nlme} can be treated as a ``correction'' to the field-independent value.
For an isotropic $s$-wave superconductor, 
the correction can be written as~\cite{2022-Makita-PRR-4-013156}:
\begin{equation}
	\label{eq:nlme}
    \lambda (T, x, B_{0}) = \lambda(T, x) \left \{ 1 + \frac{ \kappa \left ( \kappa + 2^{3/2} \right ) }{ 8 \left ( \kappa + 2^{1/2} \right )^{2} } \left [ \frac{B_{0}}{ B_{c}(T) } \right ]^{2} \right \} ,
\end{equation}
where
$B_{0}$ is the applied magnetic field,
$B_{c}(T)$ is the thermodynamic critical field,
and
$\kappa$ is the \gls{gl} parameter:
\begin{equation*}
	\kappa \equiv \frac{\lambda_\mathrm{GL}}{ \xi_{\mathrm{GL}}}, 
\end{equation*}
defined in terms of the \gls{gl} penetration depth $\lambda_\mathrm{GL}$
and
coherence length
$\xi_{\mathrm{GL}}$,
respectively.
Equivalently,
from a derivation of the \gls{gl} equations from
\gls{bcs} theory~\cite{1959-Gorkov-SPJEPT-9-1364,1960-Gorkov-SPJEPT-10-998},
$\kappa$ can be written in terms of $\lambda_{L}$, $\xi_{0}$, and $\ell$
using
(see, e.g.,~\cite{2025-Kubo-JJAP-64-018002}):
\begin{equation}
    \label{eq:kappa}
    \kappa(\ell) = \frac{\kappa_{\mathrm{clean}}}{\chi\!\bigl(a_{\mathrm{imp}}\bigr)},    
\end{equation}
where
\begin{equation}
    \label{eq:gorkov}
    \chi\!\bigl(a_{\mathrm{imp}}\bigr)
   = \frac{8}{7\,\zeta(3)}
     \sum_{n=0}^{\infty}
     \frac{1}{(2n+1)^{2}\,\bigl(2n+1 + a_{\mathrm{imp}}\bigr)} 
\end{equation}
is the Gor'kov $\chi$ function with $\zeta$ denoting the Riemann zeta function,
\begin{equation}
    \label{eq:a_imp}
    a_{\mathrm{imp}} = \frac{\pi}{2 \exp(\gamma_{E})}  \frac{\xi_{0}}{\ell} \approx 0.882 \frac{\xi_{0}}{\ell}
\end{equation}
is the impurity parameter with $\gamma_{E} = \num{0.577215}\dots$ denoting Euler's constant,
and
\begin{equation}
    \label{eq:kappa_clean}
    \kappa_{\mathrm{clean}}
  = \frac{2 \exp (\gamma_E) }{\pi} 
     \sqrt{\frac{6}{7\,\zeta(3)}} 
     \frac{\lambda_{L}}{\xi_{0}}
     \approx 0.957 \frac{\lambda_{L}}{\xi_{0}} 
\end{equation}
is the ``clean'' limit value.
Here,
we use \Cref{eq:nlme,eq:kappa,eq:gorkov,eq:a_imp,eq:kappa_clean} to account for the
scaling of $\lambda (T, x)$ in finite applied fields $B_{0}$.

\subsubsection{
	Meissner Screening Profile
	\label{sec:simulations:properties:meissner-profile}
}

Keeping with our assumption of local electrodynamics
(see \Cref{sec:simulations:properties:magnetic-penetration-depth}),
the ``classic'' London result~\cite{1935-London-PRSLA-149-71}
describing the spatial dependence of magnetic-flux-penetration
below a Meissner state superconductor's surface
can be written in a generalized form so as to account for spatial inhomogeneities
(e.g., from a depth-dependent impurity concentration).
In \gls{1d},
this generalization can be written
(in differential form)
as~\cite{1935-London-PRSLA-149-71,1981-Simon-PRB-23-4463,1986-Cave-JLTP-63-35,1994-Pambianchi-PRB-50-13659}:
\begin{widetext}
\begin{equation}
	\label{eq:london-inhomogeneous}
	\lambda^{2}(T,x,B_0) \left [ \frac{ \mathrm{d}^{2} B(x) }{ \mathrm{d}x^{2}} \right ] + 2 \lambda(T,x,B_0) \left [ \frac{ \mathrm{d} \lambda(T,x,B_0) }{ \mathrm{d}x } \right ] \left [ \frac{ \mathrm{d} B(x) }{ \mathrm{d}x } \right ] = B(x) ,
\end{equation}
\end{widetext}
where $B(x)$ is the magnetic field at depth $x$,
and
$\lambda(T,x,B_0)$ is the depth-dependent magnetic penetration
depth.\footnote{Note that when $\mathrm{d}\lambda(T,x,B_0)/\mathrm{d}x = 0$ (i.e., when $\lambda(T,x,B_0)$ is depth-independent), \Cref{eq:london-inhomogeneous} reduces to the familiar London expression~\cite{1935-London-PRSLA-149-71}, whose solution(s) follow $B(x) \propto \exp(-x / \lambda)$.}
Note that
while \Cref{eq:london-inhomogeneous} is typically used to describe systems where
screening properties are expected to vary spatially
(e.g., in proximity-coupled superconductor/normal-metal interfaces~\citenum{1981-Simon-PRB-23-4463,1994-Pambianchi-PRB-50-13659}),
analytic solutions are limited~\cite{1994-Pambianchi-PRB-50-13659},
and it generally must be solved numerically~\cite{1986-Cave-JLTP-63-35}.
Following Ref.~\cite{2024-Lechner-JAP-135-133902},
we numerically solve \Cref{eq:london-inhomogeneous}
for $B(x)$ using $\lambda(T,x,B_0)$'s spatial form derived from
the simulated oxygen defect profiles
via
\Cref{eq:mfp-rrr,eq:residual-resistivity,eq:lambda-two-fluid,eq:lambda-impurity,eq:nlme,eq:kappa,eq:gorkov,eq:a_imp,eq:kappa_clean}.

\subsubsection{
	Supercurrent Density
	\label{sec:simulations:properties:supercurrent}
}

In \gls{1d},
the supercurrent density $J(x)$ is related to $B(x)$ by its first spatial derivative:
\begin{equation}
        \label{eq:supercurrent-density}
	J(x) \equiv -\frac{1}{\mu_{0}} \frac{\mathrm{d}}{\mathrm{d}x} B(x) ,
\end{equation}
where $\mu_{0} = \qty{1.256 637 061 27 \pm 0.000 000 000 20}{\newton\per\ampere\squared}$
is the vacuum magnetic permeability~\cite{2021-Tiesinga-RMP-93-025010}.
Ultimately,
this is the most interesting quantity for \gls{srf} applications,
where there is a strong desire to:
minimize its value right at the surface
and
maximize its value deeper below the surface
(i.e., in an effort to maximize the field in
which \ch{Nb} may remain magnetic-flux-free).
Here,
we compute $J(x)$ using the $B(x)$ obtained from
numeric solutions to \Cref{eq:london-inhomogeneous}.

\subsubsection{
	Critical Supercurrent Density
	\label{sec:simulations:properties:critical-supercurrrent}
}

The critical supercurrent density $J_{c}$ defines the maximum
supercurrent that the material can withstand before magnetic vortices
penetrate its surface
(see, e.g.,~\cite{2017-Kubo-SST-30-023001}).
Within London theory, which assumes a local response,
it can be written as:
\begin{equation}
    \label{eq:critical-current-density}
    J_{c}(T, x, B_{0}) \simeq \frac{B_{c}(T)}{\mu_{0}\lambda(T, x, B_{0}) }
\end{equation}
where $B_{c}(T) \propto 1 - (T / T_{c})^{2}$ is \ch{Nb}'s thermodynamic critical field,
which at \qty{0}{\kelvin} is on the order of \qty{\sim 200}{\milli\tesla}
(see, e.g.,~\cite{1966-Finnemore-PR-149-231}).

In the context of \ch{Nb} \gls{srf} cavities with spatially inhomogeneous oxygen defect profiles
near their surface,
$J_{c}(T, x, B_{0})$ becomes depth-dependent through $\lambda(T, x, B_{0})$'s spatial
dependence. 
As $J_{c}$ defines a condition for vortex penetration,
it is thus interesting to compare it with $J(x)$ and determine
under what baking conditions, as well as applied fields,
that the inequality
\begin{equation*}
	J(x) \leq J_{c}(T, x, B_{0})
\end{equation*}
is satisfied.

\subsection{
	\label{sec:simulations:approach}
	Approach
}

With the technical aspects of the simulations described in
\Cref{sec:simulations:dissolution,sec:simulations:diffusion,sec:simulations:properties},
we now summarize their application.
In this work,
we use \Cref{eq:crank-nicolson-solution}
to approximate solutions to the reaction-diffusion system
[\Cref{eq:reaction-diffusion}].
Specifically,
we use it to predict oxygen defect profiles
in \ch{Nb} for arbitrary vacuum ``baking'' temperatures $T$ and times $t$.
As many of the best \gls{srf} cavity treatment recipes
rely on vacuum annealing
at relatively low temperatures
(i.e., $T \leq \qty{200}{\celsius}$~\cite{2004-Ciovati-JAP-96-1591,arXiv:1806.09824}),
we focus our attention on this thermal region.
Specifically,
we simulate \num{\sim 72000} different $T$ and $t$ combinations,
sampled uniformly along a ``grid'' spanning
$\qty{0.5}{\hour} \leq t \leq \qty{120}{\hour}$
and
$\qty{50}{\celsius} \leq T \leq \qty{200}{\celsius}$.
For each simulation of the oxygen defect profile,
we compute the corresponding
$\ell(x)$,
$\lambda(T, x, B_{0}$),
$B(x)$,
$J(x)$,
$J_{c}(T, x, B_{0})$,
and
$R_\mathrm{BCS}$.
In doing this, we make the following assumptions and approximations:
\begin{itemize}
\item By analogy with the plane-source models used to quantify \gls{sims} measurements
of $[\ch{O}](x)$
(see, e.g.,~\cite{2006-Ciovati-APL-89-022507,2024-Lechner-JAP-135-133902}),
we take
$[\ch{Nb2O5}]_{0} \cdot \Delta x = \qty{200}{\atp\nano\meter}$ for the oxygen within the surface oxide~\cite{2021-Lechner-APL-119-082601,2024-Lechner-JAP-135-133902},
$[\ch{O}]_{0} \cdot \Delta x = \qty{5}{\atp\nano\meter}$ for interstitial oxygen initially localized at the surface~\cite{2002-Arfaoui-JAP-91-9319,2006-Ciovati-APL-89-022507,2024-Lechner-JAP-135-133902},
$[\ch{O}]_{\mathrm{bulk}} = \qty{0.005}{\atp}$ for the ``bulk'' concentration of interstitial oxygen
(i.e., homogenized throughout the sample prior to vacuum annealing).
We also assume that the amount of \ch{NbO2} and \ch{NbO} present at $t = \qty{0}{\hour}$
is negligible
(i.e., that 
$[\ch{NbO2}]_0 \cdot \Delta x = \qty{0}{\atp\nano\meter}$
and
$[\ch{NbO}]_0 \cdot \Delta x = \qty{0}{\atp\nano\meter}$).
\item Oxygen diffusion through the oxide layer is treated as identically to inside \ch{Nb}.
For very thick oxide layers (i.e., in anodized samples) this approximation
is questionable~\cite{1985-Oechsner-TSF-124-199};
however, for ``natural'' oxide thicknesses $\lesssim \qty{5}{\nano\meter}$
it is in good accord with other
measurements~\cite{2006-Ciovati-APL-89-022507,2024-Lechner-JAP-135-133902}.
\item We take $\lambda_{L} = \qty{29}{\nano\meter}$
and $\xi_{0} = \qty{39}{\nano\meter}$ for pure \ch{Nb},
in good accord with their literature averages
(see, e.g.,~\cite{2023-McFadden-PRA-19-044018,2023-McFadden-JAP-134-163902})
and recent direct measurements using low-energy muon spin spectroscopy~\cite{2026-McFadden-PRB-113-L060508}.
\item We assume an applied field $B_{0}$ equal to \ch{Nb}'s thermodynamic critical field
$B_{c} = \qty{199.3}{\milli\tesla}$~\cite{1966-Finnemore-PR-149-231}.
While this is above \ch{Nb}'s lower critical field
$B_{c1} \approx \qty{175}{\milli\tesla}$~\cite{1966-Finnemore-PR-149-231},
it is relevant for \gls{rf} conditions such as during
\gls{srf} cavity operation,
where the best devices operate in fields above this limit
(see, e.g.,~\cite{2025-Kubo-JJAP-64-018002}).
\end{itemize}
A summary of all values used in the simulations is given in \Cref{tab:sim-params}.

\begin{table*}
\caption{
	\label{tab:sim-params}
	Summary of all adjustable parameters used in the simulation of \ch{Nb}'s surface oxide dissolution
	and diffusion in vacuum.
	The selection and justification of each parameter value are discussed in the text.
	For each quantity,
	its symbol and value are given,
	along with a brief description of the parameter's significance.
}
\begin{ruledtabular}
\begin{tabular}{l l l}
Symbol & Value & Description \\ 
\hline
$A_{0, 1}$ & \qty{9.75e8}{\per\second} & Arrhenius prefactor for the $\ch{Nb2O5} \rightarrow \ch{NbO2}$ reaction rate constant $k_1$. \\
$E_{A, 1}$ & \qty{133.7}{\kilo\joule\per\mol} & Arrhenius activation energy for $\ch{Nb2O5} \rightarrow \ch{NbO2}$ reaction rate constant $k_1$. \\
$A_{0, 2}$ & \qty{1.5e11}{\per\second} & Arrhenius prefactor for the $\ch{NbO2} \rightarrow \ch{NbO}$ reaction rate constant $k_2$.  \\
$E_{A, 2}$ & \qty{177}{\kilo\joule\per\mol} & Arrhenius activation energy for the $\ch{NbO2} \rightarrow \ch{NbO}$ reaction rate constant $k_2$.. \\
$D_{0}$ & \qty{0.59e-6}{\meter\squared\per\second} & Arrhenius prefactor for interstitial oxygen's diffusion coefficient $D$. \\
$E_{A, D}$ & \qty{109.7}{\kilo\joule\per\mol} & Arrhenius activation energy for interstitial oxygen's diffusion coefficient $D$. \\
$\sigma_e$ & \qty{3.7e-16}{\ohm\meter\squared} & Proportionality constant between \ch{Nb}'s electron mean-free-path $\ell$ and residual resistivity $\rho_e$. \\ 
$a_{\ch{O}}$ & \qty{4.5e-12}{\ohm\meter\per\ppma} & Proportionality constant between \ch{Nb}'s residual resistivity $\rho_e$ and interstitial oxygen concentration $[\ch{O}]$. \\ 
$\lambda_{L}$ & \qty{29}{\nano\meter} & \ch{Nb}'s London penetration depth. \\
$\xi_{0}$ & \qty{39}{\nano\meter} & \ch{Nb}'s \gls{bcs} coherence length. \\
$[\ch{Nb2O5}]_0 \cdot \Delta x$ & \qty{200}{\atp\nano\meter} & Surface-localized non-interstitial oxygen source at $t = \qty{0}{\hour}$. \\
$[\ch{NbO2}]_0 \cdot \Delta x$ & \qty{0}{\atp\nano\meter} & Surface-localized non-interstitial oxygen source at $t = \qty{0}{\hour}$. \\
$[\ch{NbO}]_0 \cdot \Delta x$ & \qty{0}{\atp\nano\meter} & Surface-localized non-interstitial oxygen sink at $t = \qty{0}{\hour}$. \\
$[\ch{O}]_0 \cdot \Delta x$ & \qty{5}{\atp\nano\meter} & Surface-localized interstitial oxygen concentration at $t = \qty{0}{\hour}$. \\
$[\ch{O}]_{\mathrm{bulk}} \cdot \Delta x$ & \qty{0.005}{\atp\nano\meter} & Bulk initial oxygen concentration at $t = \qty{0}{\hour}$. \\
$B_{c}$ & \qty{199.3}{\milli\tesla} & \ch{Nb}'s thermodynamic critical field. \\
$B_{0}$ & \qty{199.3}{\milli\tesla} & Applied magnetic field. \\
$T_{c}$ & \qty{9.25}{\kelvin} & \ch{Nb}'s superconducting transition temperature. \\
\end{tabular}
\end{ruledtabular}
\end{table*}

\section{
	Results
	\label{sec:results}
}

\begin{figure}
    \includegraphics[width=1.0\columnwidth]{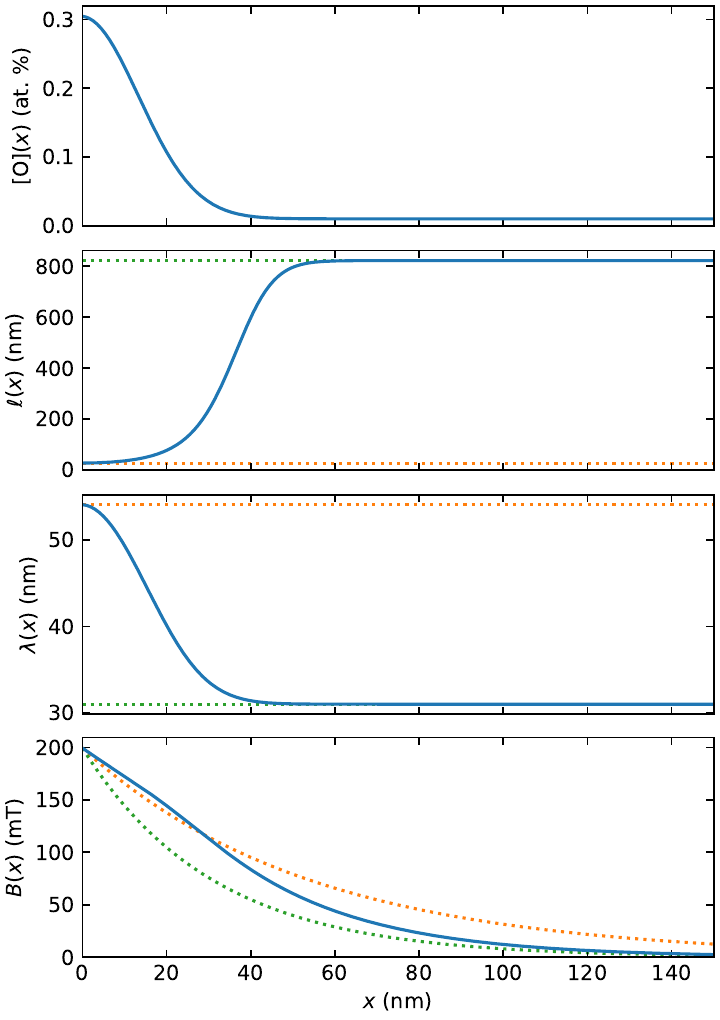}
    \caption{
        \label{fig:simulation overview}
        Simulated oxygen defect profile $[\ch{O}]$ as a function of depth $x$
        for a vacuum annealing treatment
        at \qty{125}{\celsius} for \qty{8}{\hour}.
        The simulations and derived quantities
        make use of the parameters tabulated in \Cref{tab:sim-params}.
        The spatial form of the corresponding electron mean-free-path $\ell$
        [\Cref{eq:mfp-rrr,eq:residual-resistivity}],
        magnetic penetration depth $\lambda$
        [\Cref{eq:lambda-impurity,eq:lambda-two-fluid,eq:nlme}],
        and
        Meissner screening profile $B(x)$
        [\Cref{eq:london-inhomogeneous}]
        are also shown,
        revealing surface-localized inhomogeneities resulting from $[\ch{O}](x)$.
        For comparison,
        ``clean'' and ``dirty'' limits for the latter quantities,
        identified from $\ell$'s asymptotic limits,
        are shown as dotted green and orange lines,
        respectively.
    }
\end{figure}

Typical simulation results are given in \Cref{fig:simulation overview},
where an annealing treatment of \qty{8}{\hour} at \qty{125}{\celsius}
is shown.
These parameters were selected as an illustrative example because they clearly
demonstrate the key feature of the spatially inhomogeneous current distribution discussed below.
The spatial extent of oxygen defects $[\ch{O}](x)$ resulting from the
treatment is shown,
revealing the spatial extent of doping is limited to the first \qty{\sim 40}{\nano\meter},
after which the ``bulk'' (i.e., depth-independent) value is reached.
Following \Cref{eq:mfp-rrr,eq:residual-resistivity},
this modest spatial inhomogeneity
leads to a depth-dependent electron mean-free-path $\ell$
which in turn implies a lengthened magnetic penetration depth near the surface
[\Cref{eq:lambda-impurity,eq:lambda-two-fluid}],
which returns to \ch{Nb}'s ``bulk'' value at a depth analogous to
$[\ch{O}](x)$.
This spatial variation in $\lambda(x)$ leads to distortions in the
Meissner profile obtained from \Cref{eq:london-inhomogeneous},
whose form deviates noticeably
from the (exponential) profiles predicted for
both the ``clean'' and ``dirty'' limits~\cite{2024-Lechner-JAP-135-133902,2020-Checchin-APL-117-032601}.
As a corollary of the distortions in $B(x)$,
it follows from \Cref{eq:supercurrent-density} that the
supercurrent density $J(x)$ is also deformed.

\begin{figure}
	\includegraphics[width=1.0\columnwidth]{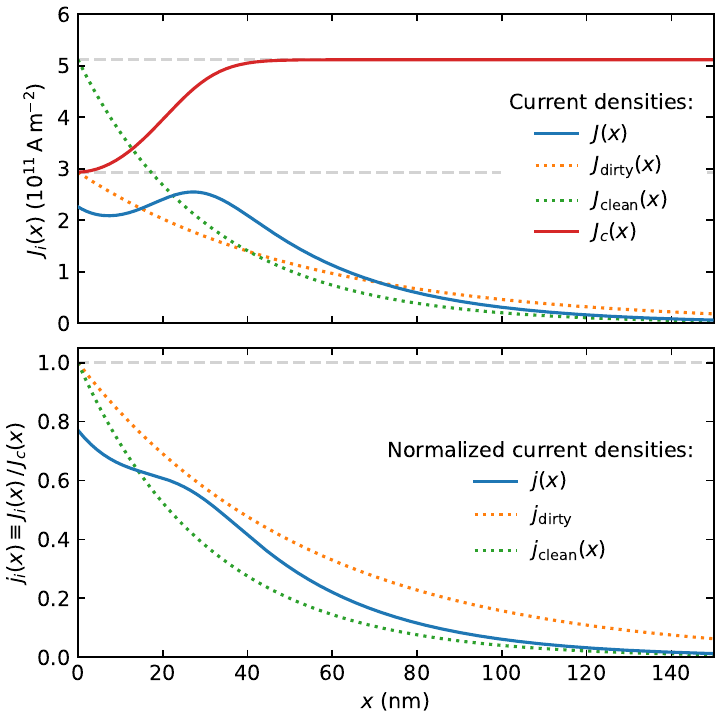}
	\caption{
		\label{fig:supercurrent}
            Supercurrent density $J$ as a function of depth $x$,
            derived from simulations of a vacuum annealing treatment
            at \qty{125}{\celsius} for \qty{8}{\hour}
            (see \Cref{fig:simulation overview})
            using \Cref{eq:supercurrent-density}.
            For comparison,
            the spatial form for both ``clean'' and ``dirty'' limits
            are shown as dotted green and orange lines,
            respectively,
            as well as \ch{Nb}'s critical current density $J_{c}$ 
            [\Cref{eq:critical-current-density}],
            which sets the limit for maintaining the Meissner state.
            Notice that this baking treatment modifies $J$
            so that its maximum coincides with the increase in $J_{c}$
            deeper below the surface.
            The fractional reduction in $J$ relative to $J_{c}$,
            represented by the quantity $j$,
            is shown on the bottom panel to emphasize the benefit of the treatment.
        }
\end{figure}

Spatial profiles for the supercurrent densities resulting from the
\qty{8}{\hour}, \qty{125}{\celsius} treatment,
derived from the $B(x)$s in \Cref{fig:simulation overview},
are shown in \Cref{fig:supercurrent}.
While the near-surface distortions in the Meissner profile from oxygen doping are subtle,
they are much more pronounced in $J(x)$,
leading to behavior that is markedly different from the (exponential)
profiles expected in the ``clean'' or ``dirty'' limits.
Instead,
$J(x)$ is significantly reduced at $x = \qty{0}{\nano\meter}$,
decreases slightly upon increasing $x$,
but then \emph{increases} to local maximum near \qty{\sim 30}{\nano\meter},
followed by exponential-like decay to zero deeper below the surface.
Consistent with simulations for closely related baking conditions~\cite{2024-Lechner-JAP-135-133902},
this highlights how the spatial form of $J(x)$ is modified by
the oxygen dissolution and diffusion process.
Explicit examples of this phenomenon
are given in \Cref{sec:additional-simulations},
including treatments
commonly applied to \gls{srf} cavitities
(see, e.g.,~\cite{2006-Visentin-PC-441-66,2004-Ciovati-JAP-96-1591,2007-Visentin-SRF-13-304}).

To contextualize the above observation,
in \Cref{fig:supercurrent}
we also plot the critical current density $J_{c}$ using \Cref{eq:critical-current-density},
which sets an upper limit for a material's ability to maintain its Meissner state.
While $J_c$ is typically depth-independent,
the spatial dependence of $\lambda(T,x,B_{0})$ imparted from doping
causes it to vary with $x$,
increasing by a factor of \num{\sim 1.5}
from near-surface to bulk depths.
The form of the $J(x)$ resulting from this baking treatment is significant;
it is reduced at the surface where $J_{c}$ smallest
and
maximized deeper within,
in close proximity to where $J_{c}$ is largest.
Notably,
the ratio $j(x) \equiv J(x) / J_{c}(x) < 1$ for all $x$,
and that the spatial form of both $J(x)$ and $j(x)$ deviate
their equivalent in the ``clean'' and ``dirty'' limits
(see \Cref{fig:supercurrent}).

\begin{figure}
	\includegraphics[width=1.0\columnwidth]{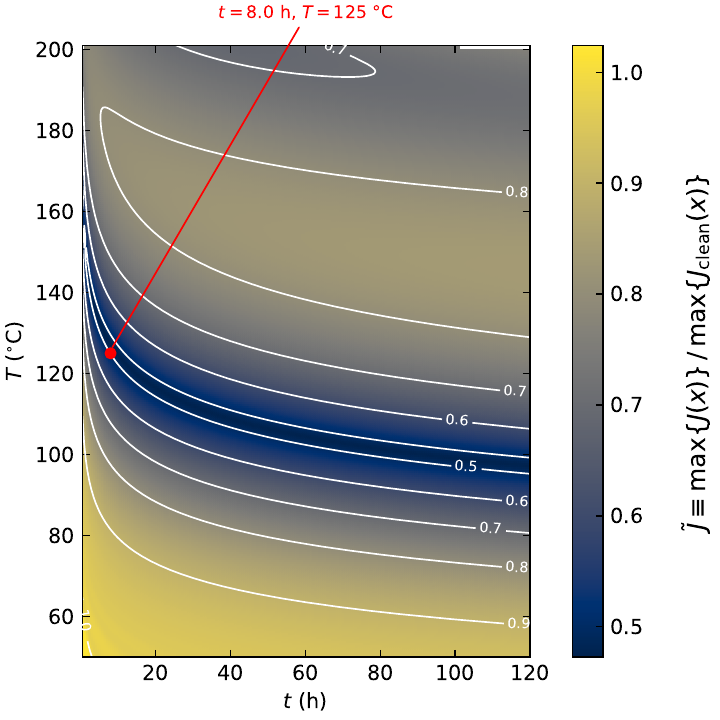}
	\caption{
		\label{fig:tilde_J}
        Ratio of the maximum supercurrent density in treated \ch{Nb} relative to its ``clean''
        value $\tilde{J}$
        [\Cref{eq:tilde_J}],
        shown for different annealing temperatures $T$ and times $t$. 
        $\tilde{J}$ varies strongly with $t$ and $T$,
        revealing a ``valley'' where it is minimized.
        The position of the \qty{8}{\hour}, \qty{125}{\celsius}
        treatment used in \Cref{fig:simulation overview,fig:supercurrent}
        is indicated,
        appearing near the bottom of the ``valley.''
	}
\end{figure}

\begin{figure}
	\includegraphics[width=1.0\columnwidth]{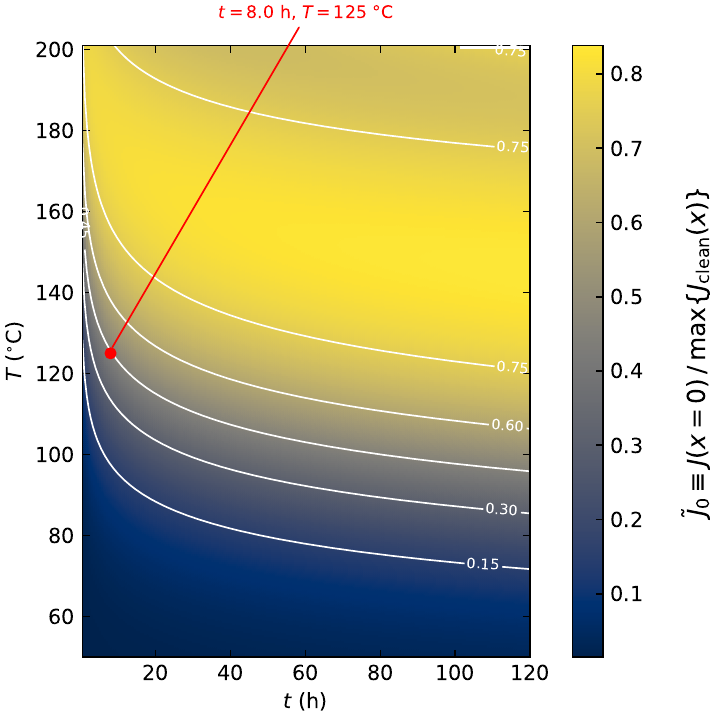}
	\caption{
		\label{fig:tilde_J_0}
        Ratio of the maximum surface supercurrent density in treated \ch{Nb} relative to its ``clean''
        value $\tilde{J}_{0}$
        [\Cref{eq:tilde_J_0}],
        shown for different annealing temperatures $T$ and times $t$.
        $\tilde{J}_{0}$ varies systematically with $t$ and $T$,
        generally increasing for ascending $t$ and $T$ values.
        The position of the \qty{8}{\hour}, \qty{125}{\celsius}
        treatment used in \Cref{fig:simulation overview,fig:supercurrent}
        is indicated.
	}
\end{figure}

\begin{figure}
	\includegraphics[width=1.0\columnwidth]{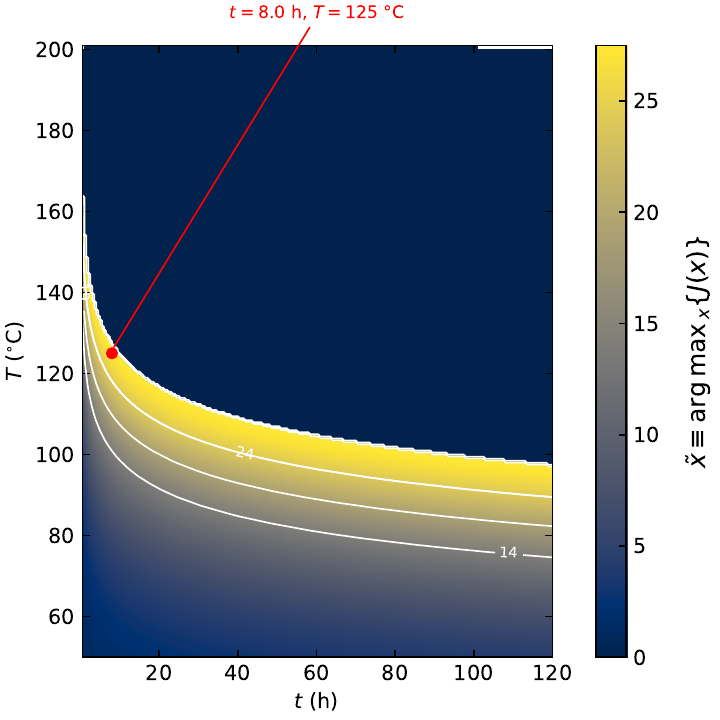}
	\caption{
		\label{fig:tilde_x}
        Position of the maximum supercurrent density in treated \ch{Nb} $\tilde{x}$
        [\Cref{eq:tilde_x}],
        shown for different annealing temperatures $T$ and times $t$. 
        $\tilde{x}$ varies strongly with $t$ and $T$,
        revealing a ``cliff'' where it is maximized.
        The position of the \qty{8}{\hour}, \qty{125}{\celsius}
        treatment used in \Cref{fig:simulation overview,fig:supercurrent}
        is indicated,
        appearing near the top of the ``cliff''.
	}
\end{figure}

Upon further inspection of \Cref{fig:supercurrent},
several additional observations can be made regarding the
impact of oxygen doping:
\begin{itemize}
\item Both $J(\qty{0}{\nano\meter})$ and $j(\qty{0}{\nano\meter})$ are reduced relative to ``clean'' \ch{Nb}.
\item Both $\max J(x)$ and $\max j(x)$ are reduced relative to ``clean'' \ch{Nb}.
\item Unlike in ``clean'' \ch{Nb}, $\max J(x)$ occurs a few nanometers below the surface rather than at the surface.
\end{itemize}
To aid in quantitatively understanding how these details are influenced by
baking conditions,
we introduce the following quantities:
\begin{equation}
    \label{eq:tilde_J}
    \tilde{J} \equiv \frac{ \max J(x)}{\max J_{\mathrm{clean}}(x) } ,
\end{equation}
\begin{equation}
    \label{eq:tilde_J_0}
    \tilde{J}_{0} \equiv \frac{J(\qty{0}{\nano\meter})}{\max J_{\mathrm{clean}}(x) } ,
\end{equation}
and
\begin{equation}
    \label{eq:tilde_x}
    \tilde{x} \equiv \arg\max_{x} \left \{  J(x)  \right \} .
\end{equation}
Plots of the temperature $T$- and time $t$-dependence of these metrics
are shown in
\Cref{fig:tilde_J,fig:tilde_J_0,fig:tilde_x}.

First we consider $\tilde{J}$,
which encapsulates the fractional reduction in the peak supercurrent density
compared to ``clean'' \ch{Nb}
(see \Cref{fig:tilde_J}).
In the clean case,
the supercurrent density is always maximized at the surface, 
whereas in treated \ch{Nb} both the peak position and magnitude
of $J(x)$ are linked to the annealing temperature and time.
We can see that,
in general,
$\tilde{J}$ is smaller for treatment temperatures
$T \lesssim \qty{150}{\kelvin}$,
with a clear ``valley'' where it is minimized.
The location of the minimum is given approximately by:
\begin{equation*}
    T(t) \approx \qty{12.3}{\celsius\per\hour^{0.196}} \cdot t^{0.196} - \qty{15.73} \cdot \ln \left ( \qty{1.23e-4}{\per\hour} \cdot t  \right ) .
\end{equation*}
This feature was also identified in Ref.~\cite{2024-Lechner-JAP-135-133902},
where the authors define the inverse of \Cref{eq:tilde_J}
as a metric of importance.
Interestingly,
a second shallower ``valley'' for $\tilde{J}$
emerges around \qty{\sim 190}{\celsius} for 
$t \gtrsim \qty{30}{\hour}$.
This likely reflects changes in the kinetics of the
dissolution reactions
(i.e., as \ch{Nb2O5} gets depleted,
the dominant source for interstitial oxygen becomes \ch{NbO2},
whose rate of dissolution is slower than the pentoxide).
Overall,
reducing $\tilde{J}$ is most readily achieved
using lower $T$ and longer $t$.

Next,
we consider the reduction of the supercurrent density right at the surface,
as summarized by the quantity $\tilde{J}_0$
(see \Cref{fig:tilde_J_0}).
In contrast to $\tilde{J}$,
the temporal and thermal evolution of $\tilde{J}_0$ is much more systematic;
low temperautures and short heating times tend to minimize surface supercurrents,
whereas
higher temperatures and longer times yield values that exceed \qty{80}{\percent} of ``clean'' \ch{Nb}.
Such an occurrence is reasonable as:
1) at low-$T$, oxygen accumulates at the surface faster than it's able to diffuse away;
and
2) at high-$T$ the rate of diffusion is high enough to (partially) homogenize the oxygen concentration profile over the London layer.
We note,
however,
that there is some deviation from the above generalization near \qty{\sim 200}{\celsius},
where the slower \ch{NbO2} dissolution rate dominates the oxygen source term in
the reaction-diffusion problem.

As alluded above,
the depth where $J(x)$ is maximized is dependent on the $t$ and $T$ of the treatment,
and we respresent this phenonmenon using the parameter $\tilde{x}$
(see \Cref{fig:tilde_x}).
The dependence of $\tilde{x}$ on $t$ and $T$ is strong,
with the most significant changes 
ocurring in the same region as the ``valley'' observed for $\tilde{J}$
in \Cref{fig:tilde_J}.
Starting at low temperautures and short treatment times,
$\tilde{x}$ increases with increasing $t$ and $T$,
quickly reaching a maximum just below \qty{\sim 30}{\nano\meter}.
The ``ridge'' defining this maximum, however,
drops off sharply to \qty{0}{\nano\meter}
upon further increases to either $t$ or $T$,
remaining there for all higher temperature and time combinations.
This suggests two things:
1) that a well-defined supercurrent peak exceeding the value at the surface
is most favourable at low temperatures and modest baking times;
and
2) there is a ``Goldilocks'' amount of baking that maximizes this quantity
(i.e., too little and the peak remains close to the surface, too much and it disappears entirely).
The ``ridge'' where $\tilde{x}$ is maximized roughly follows:
\begin{equation*}
    T(t) \approx \qty{0.8}{\celsius\per\sqrt{\hour}} \cdot \sqrt{t} - \qty{13.4}{\celsius} \cdot \ln \left (\qty{1.3e-7}{\per\hour} \cdot t \right ) .
\end{equation*}

\section{
	Discussion
	\label{sec:discussion}
}

The main effect of vacuum annealing \ch{Nb} at low temperatures is
the spatially inhomogeneous oxygen doping of its near-surface region.
Shorter baking times and lower the temperatures yield the
most abrupt changes in the dopant profile,
which in turn cause sharp changes in impurity-dependent 
quantities,
such as the electron mean-free-path $\ell$,
which have a strong influence over the metal's electrodynamics.
While changes to the Meissner response can be subtle~\cite{2024-McFadden-AIPA-14-095320},
the effect on supercurrents is significant~\cite{2024-Lechner-JAP-135-133902,2020-Checchin-APL-117-032601},
with the strongest deformations to $J(x)$ observed for low-$t$ and -$T$ combinations.
As shown in \Cref{sec:results},
both the position and magnitude of these deformations
are sensitive to the treatment conditions;
however, as illustrated above,
controlling their behaviour is highly nonlinear.

The most notable features of the $t$- and $T$-dependent simulations
are the ``cliffs'' and ``valleys'' appearing in \Cref{fig:tilde_J,fig:tilde_x} for $\tilde{x}$ and $\tilde{J}$,
whose presence suggests a local optimum.
Their absolute values rely on the kinetic balance between the accumulation of oxygen
and its propensity to disperse from concentrated regions.
As pointed out previously~\cite{2024-Lechner-JAP-135-133902},
many of the most successful ``low-$T$'' cavity treatments lie in close proximity to these regions,
suggesting their usefulness as metrics to quantify
the merit of a treatment for \gls{srf} cavity operation.
Baking temperatures above the ``cliff''/``ridge'' make changes in the dopant
profile gradual with depth, leading to a near-surface Meissner response that is virtually indistinguishable
from the ``dirty'' limit of spatially homogeneous doping.

In \gls{srf} applications,
the main importance of such preparative treatments
is to \emph{engineer} supercurrents so as to maximize cavity performance
(i.e., achieve the highest quality factor for the largest accelerating gradient).
One aspect of this is ensuring $J(x) < J_{c}(x)$;
however,
as is evident from \Cref{fig:supercurrent},
the doping treatment makes it possible
to shape $J(x)$ so as to minimize its value where $J_c(x)$
is lowest
(i.e., at surface, where the dopant concentration is highest),  
while simultaneously maximizing it deeper below the surface,
where $J_c(x)$ approaches its bulk value.
\Cref{fig:tilde_x} illustrates that,
at least when only a single temperature/time combination is considered,
there is a limit to the degree in which this can be achieved,
as evidenced by $\tilde{x}$'s abrupt drop from near its maximum to \qty{0}{\nano\meter}.

Having explored the implications of our simulations results,
we conclude the discussion with some comments on our simulation framework.
While our implementation choices (see \Cref{sec:simulations})
differ from other authors
(see, e.g.,~\cite{2024-Lechner-JAP-135-133902,2006-Ciovati-APL-89-022507,2025-Tamashevich-SST-38-045006}),
they have the advantage of generality.
While we have limited the scope of this work to
the 
$t \leq \qty{100}{\hour}$
and
$T \leq \qty{200}{\celsius}$
region,
our approach will work equally well for conditions beyond these limits,\footnote{The only caveat to this statement is that the spatial domain of the simulations must be made sufficiently large to prevent reflection of the dopant profile deep below the surface. This, however, this is easily achieved.}
including the
\qtyrange{200}{400}{\celsius} thermal region used for so-called
``mid-$T$'' treatments~\cite{2020-Posen-PRA-13-014024,2021-Ito-PTEP-2021-071G01}.
Similarly,
it is straightforward to extend the present simulation functionality
to handle \emph{sequential} treatments,
such as the famed \qty{75}{\celsius}/\qty{120}{\celsius} baking~\cite{arXiv:1806.09824}
or
other contemporary
treatment recipes~\cite{2024-Lechner-JAP-135-133902,2025-Tamashevich-SST-38-045006}.
Such a capability is closely related to
the more nuanced detail of accounting for the finite time required to heat-up
and
cool-down
a real cavity during its vacuum treatment.
While such fine-grained accounting of treatment conditions
remains relatively unexplored~\cite{2024-Lechner-JAP-135-133902},
in view of the subtleties of the baking effect,
particularly at low $T$ (see \Cref{sec:results}),
we speculate that accounting for such a fine-grained detail may be crucial towards
shaping the final $[\ch{O}](x)$ profile
and, ultimately, tailoring \ch{Nb}'s electrodynamics beyond the current state-of-the-art.

Lastly,
we note that it would be interesting to have similar simulation capabilities
for the closely related process of annealing \ch{Nb} in a nitrogenous
atmosphere,
used in so-called ``nitogen doping'' and ``nitrogen infusion'' treatments
(see, e.g.,~\cite{2020-Dhakal-PO-5-100034}).
Though the nitridation process is more complex than the surface oxide's dissolution,
progress in modeling it using an approach akin to ours has been made recently~\cite{2025-Ye-PC-628-1354616}.
To our knowledge,
there exists no modeling work on how the nitrogen dopant profiles created from such treatments
(spatially) alter \ch{Nb}'s superconducting response.
The methodology employed here to quantify pertinent features in doped \ch{Nb}'s $J(x)$
could serve as a template for such an investigation.

\section{
	Conclusions
	\label{sec:conclusions}
}

In this work,
we developed a numerical framework that combines the dissolution and diffusion
of \ch{Nb}'s native surface oxide with calculations of the resulting superconducting properties.
Using experimentally motivated kinetic and diffusion parameters,
we systematically explored treatment temperatures from \qty{50}{\celsius} to \qty{200}{\celsius}
and durations from \qty{0.5}{\hour} to \qty{120}{\hour}.
The simulations reproduce the formation of spatially inhomogeneous oxygen profiles
and
their associated modifications to the Meissner screening current distribution.
By mapping these effects across a broad parameter range,
we identify trends and treatment conditions that may provide more favourable
current distributions than conventional baking recipes.
The framework developed here provides a practical tool for evaluating
and
optimizing heat-treatment protocols and can be readily extended to time-dependent
and
multi-step processing sequences.

In the future,
it would be interesting to apply the simulation framework
to sequential vacuum heat treatments
(see, e.g.,~\cite{arXiv:1806.09824,2025-Tamashevich-SST-38-045006}),
or adapt it to the related process
of annealing \ch{Nb} in a nitrogeneous atmosphere
(see, e.g.,~\cite{2025-Ye-PC-628-1354616}).
Efforts to this end are already in progress,
and the
prospect of driving further \gls{srf} cavity development
through simulation remains strong.

\begin{acknowledgments}
	We thank E.~M.~Lechner (Thomas Jefferson National Accelerator Facility)
	for sharing some of the results in Ref.~\cite{2024-Lechner-JAP-135-133902}
	and for useful discussions.
	Financial support was provided by the \gls{nserc}.
\end{acknowledgments}

\appendix

\section{
	Reaction Rate Constants
	\label{sec:rate-constants}
}

\begin{table*}
	\centering
	\caption{
		\label{tab:kinetics}
		Summary of the Arrhenius parameters in \Cref{eq:kinetic-arrhenius}
		describing the $T$-dependence of the
		rate constants $k_{i}$ characterizing the thermal dissolution rate of
		\ch{Nb}'s native surface oxide
		[see \Cref{eq:reaction-pentoxide,eq:reaction-dioxide,eq:reaction-monoxide}].
		Here,
		$A_{0,i}$ denotes the rate process's pre-exponential factor
		and
		$E_{A,i}$ its activation energy.
		Uncertainties for each value given in parentheses,
		along with the literature source and a comment on the method of
		determination.
		Values adopted in this work are indicated by a $\star$.
	}
\begin{tabular*}{\textwidth}{l @{\extracolsep{\fill}} S S c l l}
\botrule
{Rate Constant} & {$A_{0,i}$ (\unit{\per\second})} & {$E_{A,i}$ (\unit{\kilo\joule\per\mol\per\kelvin})} & & Method & {Ref.} \\
\hline
$k_{1}$ & {$8.0 \times 10^{\num{7 \pm 0.4}} \approx \num{0.12\pm 0.14 e9}$} & 134 \pm 16 & & \gls{xps} & \cite{1990-King-TSF-192-351} \\
	& {\num{0.9 \pm 0.6 e9}} & 131 \pm 3 & & \acrshort{sims} & \cite{2021-Lechner-APL-119-082601} \\
		& {\num{0.985 \pm 0.015 e 9}} & 134.6 \pm 1.8 & & \acrshort{sims} & \cite{2024-Lechner-JAP-135-133902} \\
		& 33.96                       & 63.6 \pm 2.5 & & \gls{xps} & \cite{2024-Prudnikava-SST-37-075007} \\[1em]
	& \num{9.75 \pm 0.15 e8}                   & 133.7 \pm 1.5 & $\star$ & Weighted average of Refs.~\cite{1990-King-TSF-192-351,2021-Lechner-APL-119-082601,2024-Lechner-JAP-135-133902}  & This work \\[1em]
$k_{2}$ & {$5.4 \times 10^{\num{10.0 \pm 0.5}} \approx \num{0.10 \pm 0.18 e12}$} & 180 \pm 18 & & \gls{xps} (single-point method) & \cite{1990-King-TSF-192-351} \\
		& { $1.0 \times 10^{\num{11.0 \pm 0.4}} \approx \num{0.15 \pm 0.18 e12}$} &  177 \pm 16 & $\star$ & \gls{xps} (multi-point method) & \cite{1990-King-TSF-192-351} \\[1em]
$k_{3}$ & \sim 0 & & & \gls{xps} & \cite{1990-King-TSF-192-351} \\
\botrule
\end{tabular*}
\end{table*}

As noted in \Cref{sec:simulations:dissolution},
the temperature-dependence of the concentration of the dissolution species
[\Cref{eq:Nb2O5-concentration,eq:NbO2-concentration,eq:NbO-concentration,eq:O-concentration,eq:common-concentration}]
originates from the dissolution reaction rate constants $k_{i}$.
These $k_{i}$s are known to follow an Arrhenius law given by \Cref{eq:kinetic-arrhenius},
with several authors providing parameterizations
(see, e.g,~\cite{1990-King-TSF-192-351,2024-Prudnikava-SST-37-075007,2021-Lechner-APL-119-082601,2024-Lechner-JAP-135-133902}).
A summary of these literature values (and their uncertainties) given in \Cref{tab:kinetics}
and
we consider them in detail below.

Clearly,
there is some wide variation in the reported values
(especially for the preexponential factors),
suggesting some further discussion is warranted.
The kinetic parameters reported in Ref.~\cite{2024-Prudnikava-SST-37-075007}
are the obvious outliers
and
a careful read of the manuscript suggests that their data is the least trustworthy
(i.e.,
they cover only a limited temperature range,
they took a very limited number of measurements,
and
they report issues with promptly achieving temperature
stability).\footnote{Note that in Ref.~\cite{2024-Prudnikava-SST-37-075007} they analyze their data under the assumptions of both first- \emph{and} second-order reactions, which they differentiate by using the $k_{1}$ and $k_{2}$ notation, respectively. As discussed in Ref.~\cite{1990-King-TSF-192-351}, these reactions are \emph{first-order}, so the motivation for doing this is unclear. Consequently, we rule out the merit of Ref.~\cite{2024-Prudnikava-SST-37-075007}'s $k_{2}$ values on the basis that they are inconsistent with existing kinetic models for the dissolution process.}
Moreover,
their measurements are likely in the low-$T$ limit of what is reasonable for
accurately following the dissolution process
(cf.~\cite{1990-King-TSF-192-351}).
Interestingly,
the highest $k_{1}$ in~\cite{2024-Prudnikava-SST-37-075007} is
consistent with the data in~\cite{1990-King-TSF-192-351},
with the estimates at lower-$T$
deviating from a common Arrhenius trend.
This may suggest a systematic ``floor'' for the measured values.
As no uncertainties are quoted
and they are the outlier among the other reported
values~\cite{1990-King-TSF-192-351,2021-Lechner-APL-119-082601,2024-Lechner-JAP-135-133902},
we do not consider them further.

Noting the similarity of the results reported in Refs.~\cite{1990-King-TSF-192-351,2021-Lechner-APL-119-082601,2024-Lechner-JAP-135-133902},
we adopt the following values for the $A_{0,i}$s and $E_{A,i}$s
describing the $k_{1}$ and $k_{2}$ rate constants.
For $k_{1}$,
we use values computed from their \emph{weighted average},
giving
$A_{0,1} = \qty{9.75e8}{\per\second}$ and $E_{A,1} = \qty{133.7}{\kilo\joule\per\mol}$ for $k_{1}$.
For $k_{2}$,
where there are less measurements available,
we follow the recommendation in Ref.~\cite{1990-King-TSF-192-351}
and
use the Arrhenius obtained from their ``multi-point method'':
$A_{0,2} = \qty{1.5e11}{\per\second}$ and $E_{A,2} = \qty{177}{\kilo\joule\per\mol}$ for $k_{2}$.
Note that in assigning these values,
we have transformed the results in Ref.~\cite{1990-King-TSF-192-351}
to a form where the uncertainties are no longer in the exponent
(using Monte Carlo error propagation).
The adopted $A_{0,i}$ and $E_{A,i}$ values are indicated in \Cref{tab:kinetics}.

\section{
	Interstitial Oxygen Diffusivity in Nb
	\label{sec:oxygen-diffusion}
}

Here we discuss details pertaining the diffusion of oxygen in \ch{Nb}.
As noted in \Cref{sec:simulations:diffusion},
the diffusion coefficient $D$ for interstitial oxygen in \ch{Nb} follows
an Arrhenius temperature dependence,
characterized by a prefactor $D_{0}$ and activation energy $E_{A,D}$
[see \Cref{eq:diffusion-arrhenius}].
Values for these quantities have been determined by many authors
(see, e.g.,~\cite{1959-Powers-JAP-30-514,1966-Giabala-AM-14-1095,1977-Kirchheim-ZM-68-97,1977-Perkins-AM-25-1221,1977-Boratto-MTA-8-1233,1977-Boratto-SM-11-709,1979-Kirchheim-AM-27-869,1979-Lauf-AM-27-1157,1979-Farraro-MSE-39-47,1980-Boratto-MSE-43-97}),
with most of the results summarized in~\cite{1990-LeClaire-LBIII-26-471}.
The \emph{best} parameter estimates come from Refs.~\cite{1977-Boratto-SM-11-709,1980-Boratto-MSE-43-97},
where it was shown that a single Arrhenius expression is sufficient to describe
oxygen's diffusivity from \qtyrange{23}{1545}{\celsius}.
One obtains slightly different values depending on if the diffusion ``pathway''
goes through an octahedral or tetrahedral interstitial site in \ch{Nb}'s \gls{bcc}
crystal lattice
(see, e.g.,~\cite{1965-Beshers-JAP-36-290});
however, they two situations yield very similar results.
For migration via octahedral sites,
$D_{0} = \qty{0.59 \pm 0.09 e-6}{\meter\squared\per\second}$
and
$E_{A,D} = \qty{109.7 \pm 0.6}{\kilo\joule\per\mol\per\kelvin}$;
for tetrahedral sites,
the term are
$D_{0} = \qty{0.69 \pm 0.12 e-6}{\meter\squared\per\second}$
and
$E_{A,D} = \qty{112.3 \pm 0.7}{\kilo\joule\per\mol\per\kelvin}$.
Though more recent microscopic measurements suggest there may be ambiguity in the
atomistic details of the mass transport~\cite{1994-Michel-AMM-42-3409},
based on the implicit~\cite{1990-LeClaire-LBIII-26-471}
and
explicit~\cite{1979-Kirchheim-AM-27-869}
recommendations that the assumption of an octahedral ``pathway'' gives the
best description of diffusivity over both short- and long-range distances,
we adopt the octahedral kinetic parameters for our simulations.\footnote{Interestingly, the recommended Arrhenius parameters are somewhat smaller than those inferred from recent \gls{sims}
measurements of ``mid-$T$'' baked \ch{Nb}~\cite{2021-Lechner-APL-119-082601,2024-Lechner-JAP-135-133902}; however, this difference is understandable, given
the relatively small $T$-range used in the measurements
(i.e., \qtyrange{300}{600}{\kelvin}~\cite{2024-Lechner-JAP-135-133902} vs.\
\qtyrange{23}{1545}{\celsius}~\cite{1977-Boratto-SM-11-709,1977-Boratto-MTA-8-1233}).}

\section{
	Crank-Nicolson Method
	\label{sec:crank-nicolson}
}

The \gls{cn} method is a finite difference ``stencil'' scheme developed in the 1940s
to solve the heat equation and related partial differential equations numerically~\cite{1947-Crank-MPCPS-43-50}.
As alluded in \Cref{sec:simulations:diffusion},
the method uses \emph{discretization}
as a means of finding an approximation for the concentration $[\ch{O}](x,t)$
that solves \Cref{eq:reaction-diffusion}.
As this concentration is really a \gls{2d} quantity,
discretization is taken over both the spatial variable $x$
and
the temporal variable $t$,
forming a \gls{2d} regular grid with
$N_{t}$ time points separated by $\Delta t$ time steps along one axis,
and
$N_{x}$ spatial points separated by $\Delta x$ position steps.
The goal then is to approximate the (unknown) $[\ch{O}](x,t)$ in these discrete grid points
such that:
\begin{equation*}
	O(n_{x} \Delta x, n_{t} \Delta t) \approx [\ch{O}](x,t) ,
\end{equation*}
where $O$ is the numeric approximation of $[\ch{O}]$,
$n_{i} \Delta i = i$,
and
$n_{i} = 0, 1, 2, \dots, N_{i} - 2, N_{i} - 1$.
From now on,
we use the shorthand notation
$O_{n_{x}}^{n_{t}}$ to refer to the solution at grid point
$(n_{x} \Delta x, n_{t} \Delta t)$.
Similarly,
in the following we assume Neumann boundary conditions
(i.e., $\partial [\ch{O}](x,t) / \partial x |_{x = 0, L} = 0$,
where
where $x \in [0, L]$ and $L$ is the length of the
spatial domain).\footnote{For the purpose of simulating the dissolution/diffusion of \ch{Nb}'s surface oxide, $L$ should always be chosen to be larger than the spatial extent of the produced oxygen profile (i.e., otherwise some of the oxygen atoms will be ``reflected'' at this boundary and artificially inflate the profile's ``baseline'').}

To approximate the time-derivative on the left-hand side of \Cref{eq:reaction-diffusion},
we use values of $O$ at two specific grid points:
\begin{equation}
	\label{eq:num-t-deriv}
	 \left . \frac{\partial}{\partial t} [\ch{O}](x,t) \right |_{x, t} \approx \frac{ O_{n_{x}}^{n_{t}+1} - O_{n_{x}}^{n_{t}}}{\Delta t} .
\end{equation}
Clearly,
this is a (forward) finite difference scheme (i.e., stencil) that we wish to superimpose on our
$(n_{x}, n_{t})$ grid.
Examining the right-hand side of \Cref{eq:reaction-diffusion},
we must also numerically approximate the 2\textsuperscript{nd} spatial derivative of $O$
(i.e., the \gls{1d} Laplace operator $\nabla^{2} \equiv \nabla \cdot \nabla = \partial^{2} / \partial x^{2}$):
\begin{align}
	\label{eq:laplacian-def}
	\left . \frac{\partial [\ch{O}](x,t)^{2}}{\partial x^{2}} \right |_{x, t} &\approx \frac{ \frac{ O_{n_{x}+1}^{n_{t}} - O_{n_{x}}^{n_{t}} }{\Delta x} - \frac{ O_{n_{x}}^{n_{t}} - O_{n_{x} - 1}^{n_{t}} }{\Delta x} }{\Delta x} \\
	\label{eq:laplacian}
	&\approx \frac{ O_{n_{x} + 1}^{n_{t}} - 2 O_{n_{x}}^{n_{t}} + O_{n_{x} - 1}^{n_{t}} }{ (\Delta x)^{2} } .
\end{align}
While the application of \Cref{eq:laplacian} is clear,
were it used directly we would arrive at the (explicit) \gls{ftcs} method for solving the diffusion equation.
Alternatively,
one could write \Cref{eq:laplacian} where time points $n_{t} + 1$ were used in place of those at $n_{t}$
and
we would obtain the (implicit) \gls{btcs} approach.
Instead,
the \gls{cn} approach uses an (implicit) \gls{ctcs} method,
wherein the \emph{average} of the Laplacians at $n_{t}$ and $n_{t} + 1$ are used:
\begin{widetext}
\begin{equation}
	\label{eq:num-x-deriv2}
	 \left . \frac{\partial [\ch{O}](x,t)^{2}}{\partial x^{2}} \right |_{x, t} \approx \frac{1}{ 2 (\Delta x)^{2} } \left ( O_{n_{x} + 1}^{n_{t}} - 2 O_{n_{x}}^{n_{t}} + O_{n_{x} - 1}^{n_{t}} + O_{n_{x} + 1}^{n_{t} + 1} - 2 O_{n_{x} }^{n_{t} + 1} + O_{n_{x} - 1}^{n_{t} + 1} \right ) .
\end{equation}
\end{widetext}
In essence,
\Cref{eq:num-x-deriv2} is like taking the 2\textsuperscript{nd} derivative at $n_{t} + 1/2$
and
the inclusion of the two time points in conjunction with \Cref{eq:num-t-deriv} makes it clear
how the \gls{cn} ``stencil'' arises.
Lastly,
we need to account for the source term $q(x, t)$ in \Cref{eq:reaction-diffusion},
which is defined in \Cref{eq:source}.
In this case,
our approximation simply
amounts to its evaluation at discrete temporal/spatial points:
\begin{equation}
	\label{eq:num-source}
	q(x, t) \approx Q(n_{x} \Delta x, n_{t} \Delta t) .
\end{equation}

With the individual approximations defined above,
we may now write out the \gls{cn} scheme in full.
Combining \Cref{eq:num-t-deriv,eq:num-x-deriv2,eq:num-source},
\Cref{eq:reaction-diffusion} can be written approximately as:
\begin{widetext}
\begin{equation}
	\label{eq:diffusion-general-approx}
	\frac{ O_{n_{x}}^{n_{t}+1} - O_{n_{x}}^{n_{t}}}{\Delta t} = \frac{D}{2 (\Delta x)^{2}} \left ( O_{n_{x} + 1}^{n_{t}} - 2 O_{n_{x}}^{n_{t}} + O_{n_{x} - 1}^{n_{t}} + O_{n_{x} + 1}^{n_{t} + 1} - 2 O_{n_{x} }^{n_{t} + 1} + O_{n_{x} - 1}^{n_{t} + 1} \right ) + Q_{n_{x}}^{n_{t}} ,
\end{equation}
\end{widetext}
where $D$ is the diffusion coefficient.
Noting that \Cref{eq:diffusion-general-approx} contains both $n_{t}$ and $n_{t}+1$ time points,
it is beneficial to re-order the expression so as to isolate these terms on either side of the equation:
\begin{widetext}
\begin{align}
	\label{eq:diffusion-general-approx-reorder}
    -\sigma O_{n_{x}-1}^{n_{t}+1} + (1 + 2 \sigma) O_{n_{x}}^{n_{t}+1} - \sigma O_{n_{x}+1}^{n_{t}+1} = \sigma O_{n_{x}-1}^{n_{t}} + (1 - 2 \sigma)O_{n_{x}}^{n_{t}} + \sigma O_{n_{x}+1}^{n_{t}} + \Delta t Q_{n_{x}}^{n_{t}} ,
\end{align}
\end{widetext}
where $n_{x} = 2, 3, \dots, N_{x} - 3, N_{x} - 2$.
Here in \Cref{eq:diffusion-general-approx-reorder} we have introduced the (dimensionless) term:
\begin{equation*}
	\sigma \equiv \frac{D \Delta t}{ 2 (\Delta x)^{2} } = \frac{r}{2} ,
\end{equation*}
where $r \equiv D \Delta t / (\Delta x)^{2}$ is the von~Neumann stability term~\cite{1947-Crank-MPCPS-43-50,1950-Charney-T-2-237},
so as to express the problem more compactly.\footnote{Note that the \gls{cn} method has the special property that it is \emph{unconditionally} stable for all $\sigma$; however, when $r > 1/2$ (i.e., when the grid spacing ``spread'' $(\Delta x)^{2}/\Delta t$ is less than or comparable to the diffusivity $D$), oscillatory components in the approximate solutions $O$ are possible.}
It is also worth noting that both sides of the expression are dimensionally correct,
each having units of concentration.

Before proceeding,
a commentary on some of the technical details of \Cref{eq:diffusion-general-approx-reorder}
is required.
As defined above,
this expression is valid for all non-boundary spatial points
$n_{x} = 1, 2, \dots, N_{x} - 3, N_{i} - 2$;
however, problems are encountered for indices $n_{x} = 0$ and $n_{x} = N_{x} - 1$
(i.e., the values $O_{-1}^{n_{t}}$ and $O_{N_{x}}^{n_{t}}$ that arise at these boundaries
both lie \emph{outside} of our spatial-temporal grid).
Fortunately,
by imposing the Neumann boundary conditions,
it is straightforward to work out what these (literal) edge cases should be.
Inspecting the definition for the numeric approximation of the Laplacian in \Cref{eq:laplacian-def},
we see that $\partial c / \partial x = 0$ is only satisfied at the spatial boundaries when:
\begin{align*}
	n_{x} &= 0:        & O_{-1}^{n_{t}} &= O_{0}^{n_{t}} & O_{-1}^{n_{t}+1} &= O_{0}^{n_{t}+1} \\
	n_{x} &= N_{x} -1: & O_{N_{x}}^{n_{t}} &= O_{N_{x}-1}^{n_{t}} & O_{N_{x}}^{n_{t}+1} &= O_{N_{x}-1}^{n_{t}+1} \\
\end{align*}
Upon substituting these edge-case relations into \Cref{eq:diffusion-general-approx-reorder} we obtain:
\begin{widetext}
\begin{align}
	\label{eq:diffusion-general-approx-reorder-0}
	(1 + \sigma) O_{n_{x}}^{n_{t}+1} - \sigma O_{n_{x}+1}^{n_{t}+1} &= (1 - \sigma)O_{n_{x}}^{n_{t}} + \sigma O_{n_{x}+1}^{n_{t}} + \Delta t Q_{n_{x}}^{n_{t}} ,  &  n_{x} &= 0 , \\
	\label{eq:diffusion-general-approx-reorder-N-1}
	-\sigma O_{n_{x}-1}^{n_{t}+1} + (1 + \sigma) O_{n_{x}}^{n_{t}+1}  &= \sigma O_{n_{x}-1}^{n_{t}} + (1 - \sigma)O_{n_{x}}^{n_{t}} + \Delta t Q_{n_{x}}^{n_{t}} ,  &  n_{x} &= N_{x} - 1 . 
\end{align}
\end{widetext}
Together,
\Cref{eq:diffusion-general-approx-reorder,eq:diffusion-general-approx-reorder-0,eq:diffusion-general-approx-reorder-N-1}
cover the full range of the \gls{cn} spatial-temporal grid.

To put the \gls{cn} scheme into action,
one additional ``transformation'' is required.
It is instructive to consider a \emph{full} set of spatial solutions
\begin{equation*}
\{ O_{0}^{n_{t}}, O_{1}^{n_{t}}, O_{2}^{n_{t}}, \dots, O_{N_{x}-2}^{n_{t}}, O_{N_{x}-1}^{n_{t}} \}
\end{equation*}
at a fixed point in time $t = n_{t} \Delta t$
as a single mathematical entity.
Naturally,
this set is easily represented as a column vector:
\begin{equation}
	\label{eq:concentration-vector}
	\mathbf{O}^{n_{t}} = \begin{bmatrix}
		O_{0}^{n_{t}} \\
		O_{1}^{n_{t}} \\
		\vdots \\
		O_{N_{x}-2}^{n_{t}} \\
		O_{N_{x}-1}^{n_{t}}
	  \end{bmatrix} ,
\end{equation}
where $\mathbf{O}^{n_{t}}$'s elements correspond to different spatial points along its grid.
The advantage of using this notation is that
the derivative approximations used above can be written as \emph{matricies},
culminating in a compact representation of the system of equations
given by \Cref{eq:diffusion-general-approx-reorder,eq:diffusion-general-approx-reorder-0,eq:diffusion-general-approx-reorder-N-1}.
Explicitly,
they
can be expressed as:
\begin{equation}
	\label{eq:diffusion-general-cn-matrix}
	\mathsf{A} \mathbf{O}^{n_{t}+1} = \mathsf{B} \mathbf{O}^{n_{t}} + \Delta t \mathbf{Q}^{n_{t}} ,
\end{equation}
where
\begin{widetext}
\begin{equation}
	\label{eq:A-matrix}
	\mathsf{A} \equiv
	\begin{bmatrix}
		1 + \sigma &   - \sigma & 0 &  \cdots & 0 & 0 & 0 \\
		  - \sigma & 1 + 2\sigma & -\sigma &  \cdots & 0 & 0 & 0 \\
		         0 &    -\sigma & 1 + 2 \sigma & \cdots & 0 & 0 & 0  \\
		         \vdots &      \vdots     & \vdots  & \ddots & \vdots & \vdots & \vdots \\
		0  & 0 & 0 &  \cdots & 1 + 2\sigma & -\sigma & 0 \\
		0  & 0 & 0 &  \cdots & -\sigma & 1 + 2\sigma & -\sigma \\
		0  & 0 & 0 &  \cdots & 0 & -\sigma & 1 + \sigma \\
	\end{bmatrix}
\end{equation}
and
\begin{equation}
	\label{eq:B-matrix}
	\mathsf{B} \equiv
	\begin{bmatrix}
		1 - \sigma &    \sigma & 0 & \cdots & 0 & 0 & 0 \\
		   \sigma & 1 - 2\sigma & \sigma & \cdots & 0 & 0 & 0 \\
		         0 &    \sigma & 1 - 2 \sigma & \cdots & 0 & 0 & 0 \\
		         \vdots &      \vdots     & \vdots & \ddots & \vdots & \vdots & \vdots \\
		0  & 0 & 0 & \cdots & 1 - 2\sigma & \sigma & 0 \\
		0  & 0 & 0 & \cdots & \sigma & 1 - 2\sigma & \sigma \\
		0  & 0 & 0 & \cdots & 0 & \sigma & 1 - \sigma \\
	\end{bmatrix}
\end{equation}
\end{widetext}
describe the transfer of concentration $\mathbf{O}^{i}$ between spatial coordinates
upon moving from the $n_{t}$ to the $n_{t + 1}$ time step,
$\mathbf{O}^{i}$ is given by \Cref{eq:concentration-vector},
and
\begin{equation}
	\label{eq:source-vector}
	\mathbf{Q}^{n_{t}} \equiv \begin{bmatrix}
		Q_{0}^{n_{t}} \\
		Q_{1}^{n_{t}} \\
		\vdots \\
		Q_{N_{x}-2}^{n_{t}} \\
		Q_{N_{x}-1}^{n_{t}}
	\end{bmatrix}
\end{equation}
denotes the source term in column vector form.

Since the system of equations in \Cref{eq:diffusion-general-cn-matrix}
always starts with a well-defined initial condition
(i.e., knowledge of $\mathbf{C}^{0}$ is a \emph{required} starting point),
the expressions have only one unknown
(i.e., the vector $\mathbf{O}^{n_{t}+1}$).
Therefore,
the solution at each \emph{proceeding} time step is given
by inverting $\mathsf{A}$ and re-arranging for $\mathbf{O}^{n_{t}+1}$,
which yields the expression for \Cref{eq:crank-nicolson-solution}
introduced in \Cref{sec:simulations:diffusion}.

To summarize,
the numeric approximations employed by the \gls{cn} method~\cite{1947-Crank-MPCPS-43-50,1975-Crank-TMOD-2}
culminate in a set of (linear) expressions
[\Cref{eq:diffusion-general-approx-reorder,eq:diffusion-general-approx-reorder-0,eq:diffusion-general-approx-reorder-N-1}]
that may be written compactly in matrix form
[\Cref{eq:diffusion-general-cn-matrix}]
and
solved using linear algebra techniques
[\Cref{eq:crank-nicolson-solution}].
The approach provides a
a clear and concise means of numerically evolving
an initial concentration ``profile'' according to
\Cref{eq:reaction-diffusion}.

\section{
    Additional Simulation Results
    \label{sec:additional-simulations}
}

\newcommand\figtextfrac{0.245}

\begin{figure*}[hp]
    \centering
    \includegraphics[width=\figtextfrac\textwidth]{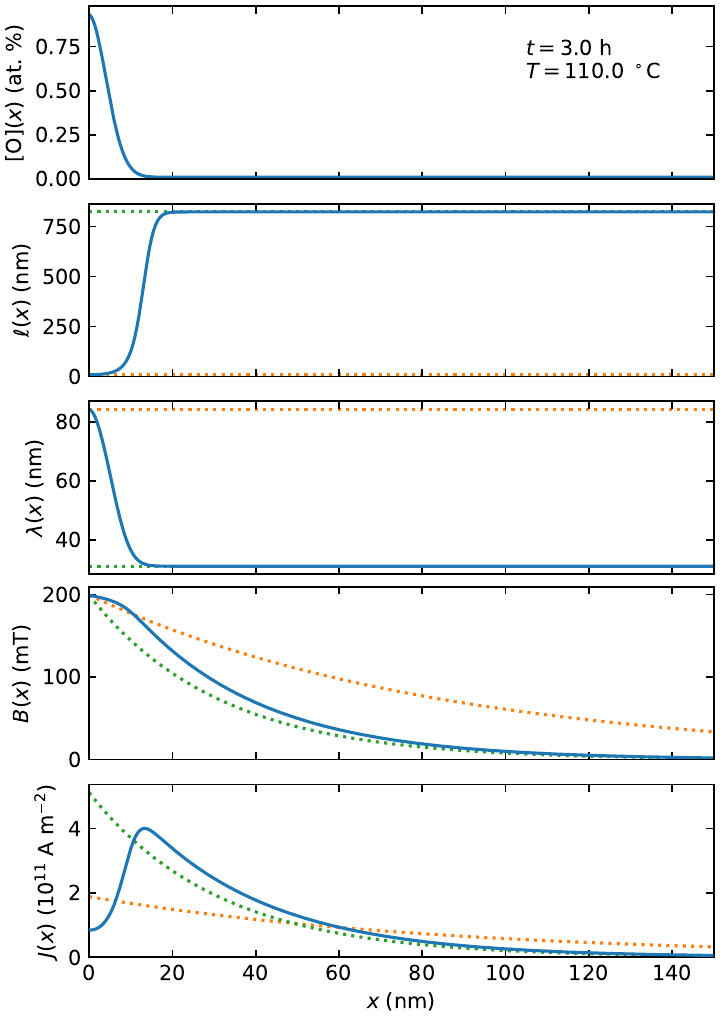} \hfill
    \includegraphics[width=\figtextfrac\textwidth]{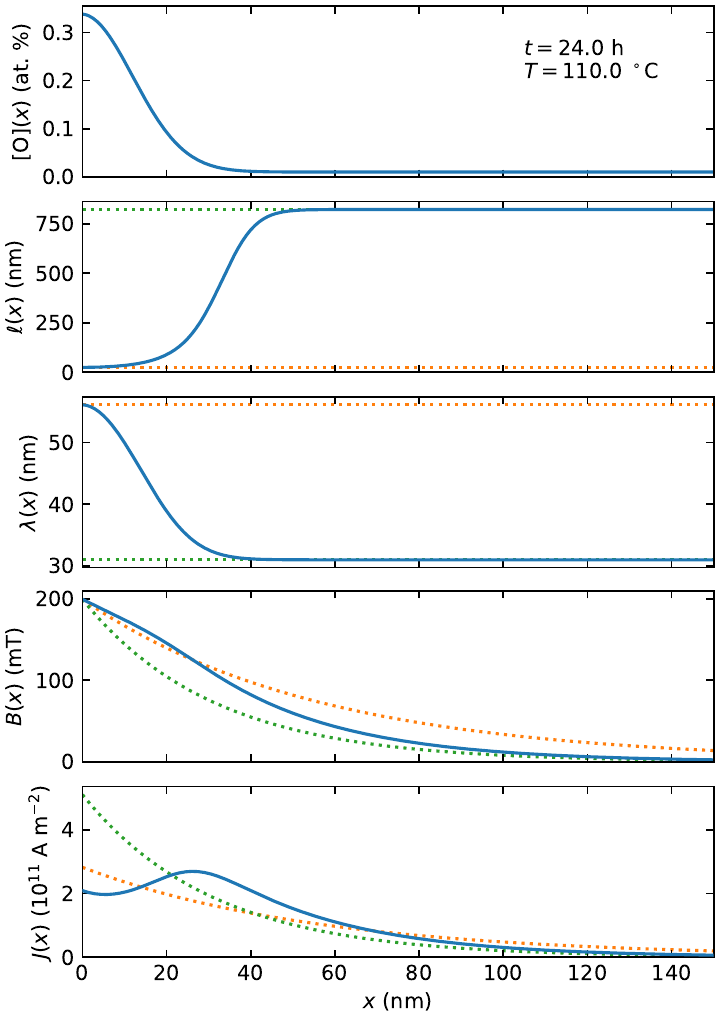} \hfill
    \includegraphics[width=\figtextfrac\textwidth]{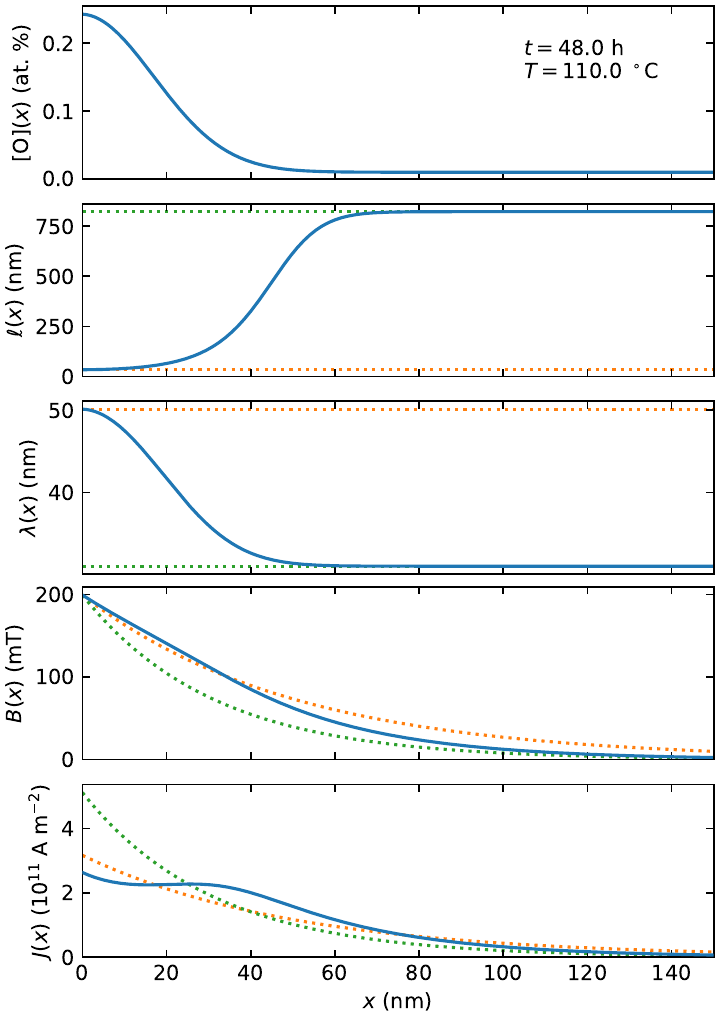} \hfill
    \includegraphics[width=\figtextfrac\textwidth]{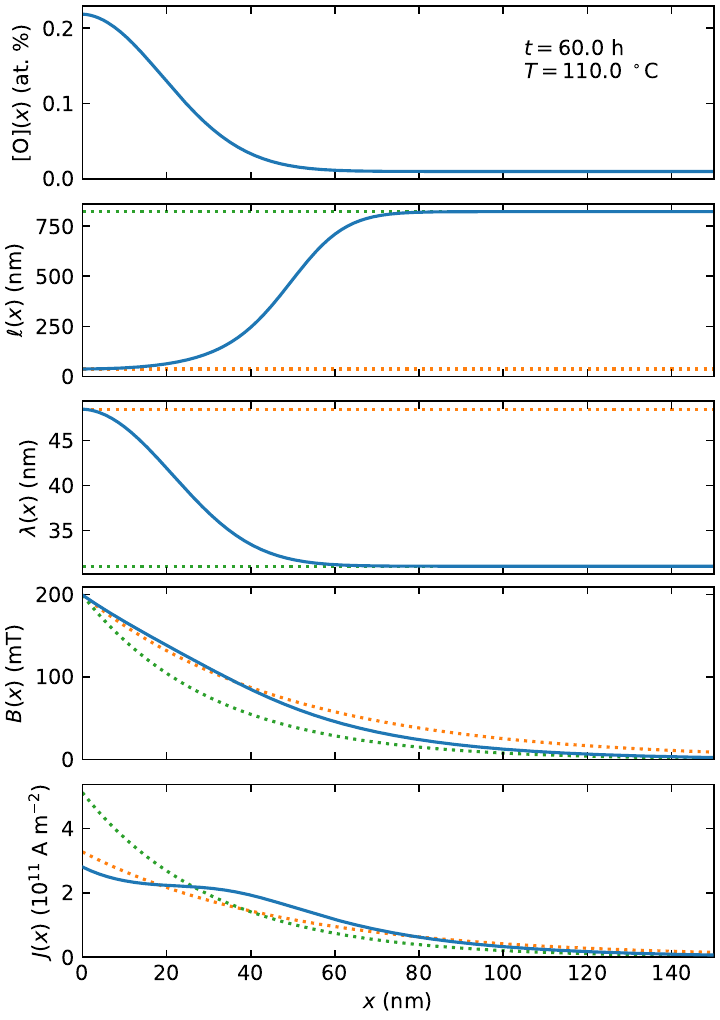}
    \caption{
        \label{fig:sim-110C}
        Simulated oxygen defect profile $[\ch{O}]$ as a function of depth $x$
        for several vacuum annealing treatment times $t$
        (indicated in each plot's top panel)
        at \qty{110}{\celsius}.
        The simulations and derived quantities
        make use of the parameters tabulated in \Cref{tab:sim-params}.
        The spatial form of the corresponding electron mean-free-path $\ell$
        [\Cref{eq:mfp-rrr,eq:residual-resistivity}],
        magnetic penetration depth $\lambda$
        [\Cref{eq:lambda-impurity,eq:lambda-two-fluid,eq:nlme}],
        Meissner screening profile $B(x)$
        [\Cref{eq:london-inhomogeneous}],
        and
        supercurrent density $J(x)$
        [\Cref{eq:supercurrent-density}],
        are also shown,
        revealing surface-localized inhomogeneities resulting from $[\ch{O}](x)$.
        For comparison,
        ``clean'' and ``dirty'' limits for the latter quantities,
        identified from $\ell$'s asymptotic limits,
        are shown as dotted green and orange lines,
        respectively.
    }
\end{figure*}

\begin{figure*}[hp]
    \centering
    \includegraphics[width=\figtextfrac\textwidth]{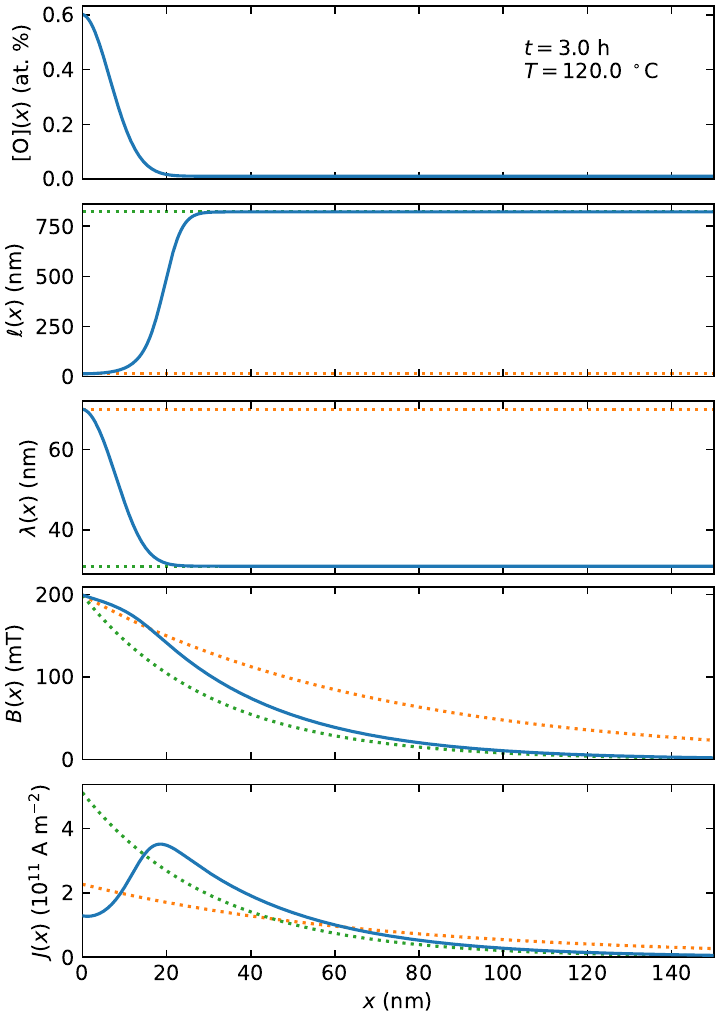} \hfill
    \includegraphics[width=\figtextfrac\textwidth]{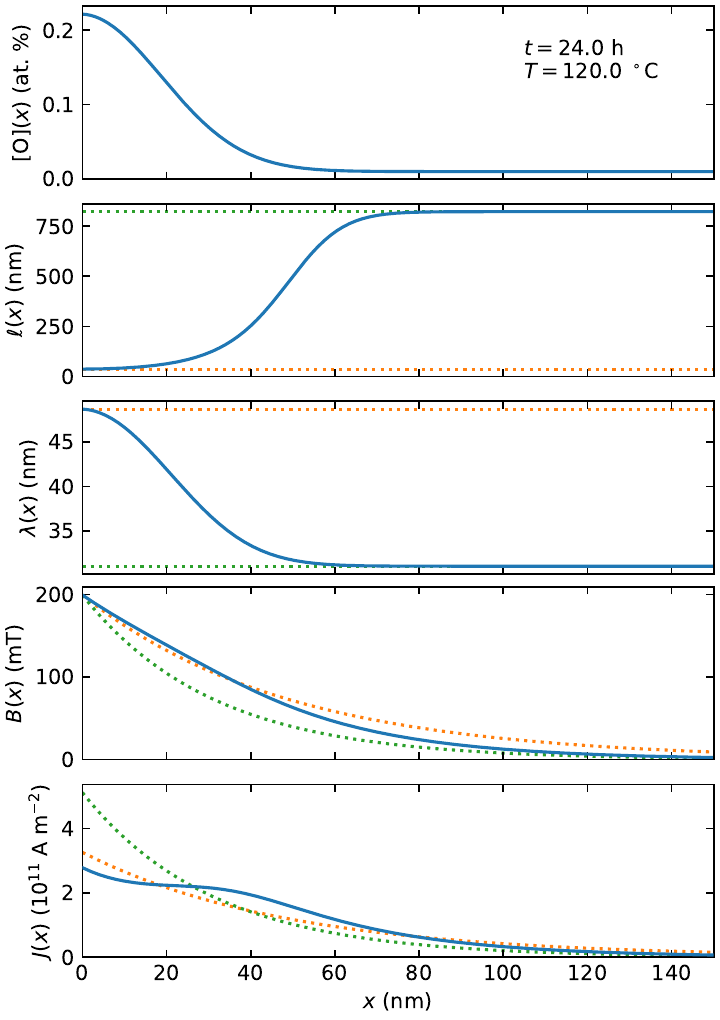} \hfill
    \includegraphics[width=\figtextfrac\textwidth]{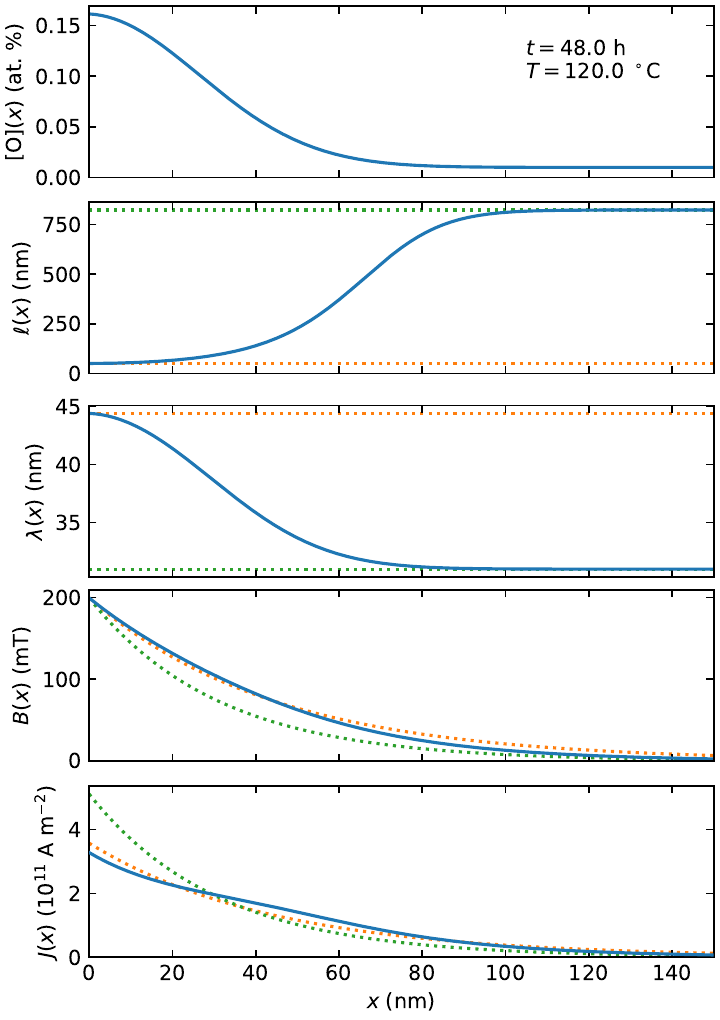} \hfill
    \includegraphics[width=\figtextfrac\textwidth]{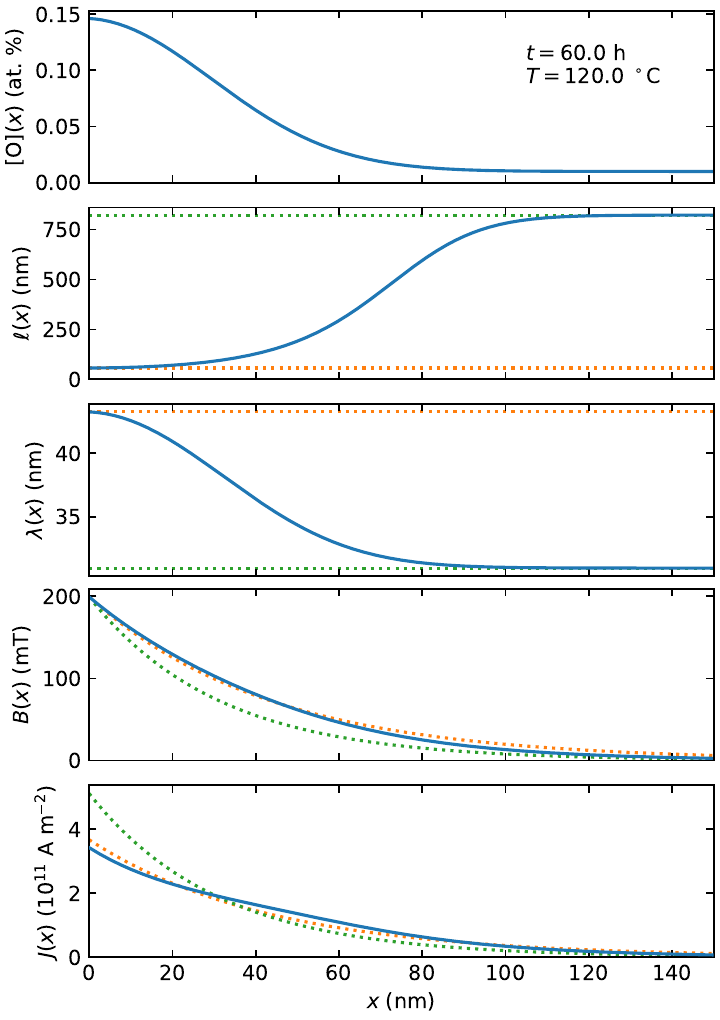}
    \caption{
        \label{fig:sim-120C}
        Simulated oxygen defect profile $[\ch{O}]$ as a function of depth $x$
        for several vacuum annealing treatment times $t$
        (indicated in each plot's top panel)
        at \qty{120}{\celsius}.
        The simulations and derived quantities
        make use of the parameters tabulated in \Cref{tab:sim-params}.
        The spatial form of the corresponding electron mean-free-path $\ell$
        [\Cref{eq:mfp-rrr,eq:residual-resistivity}],
        magnetic penetration depth $\lambda$
        [\Cref{eq:lambda-impurity,eq:lambda-two-fluid,eq:nlme}],
        Meissner screening profile $B(x)$
        [\Cref{eq:london-inhomogeneous}],
        and
        supercurrent density $J(x)$
        [\Cref{eq:supercurrent-density}],
        are also shown,
        revealing surface-localized inhomogeneities resulting from $[\ch{O}](x)$.
        For comparison,
        ``clean'' and ``dirty'' limits for the latter quantities,
        identified from $\ell$'s asymptotic limits,
        are shown as dotted green and orange lines,
        respectively.
    }
\end{figure*}

\begin{figure*}[hp]
    \centering
    \includegraphics[width=\figtextfrac\textwidth]{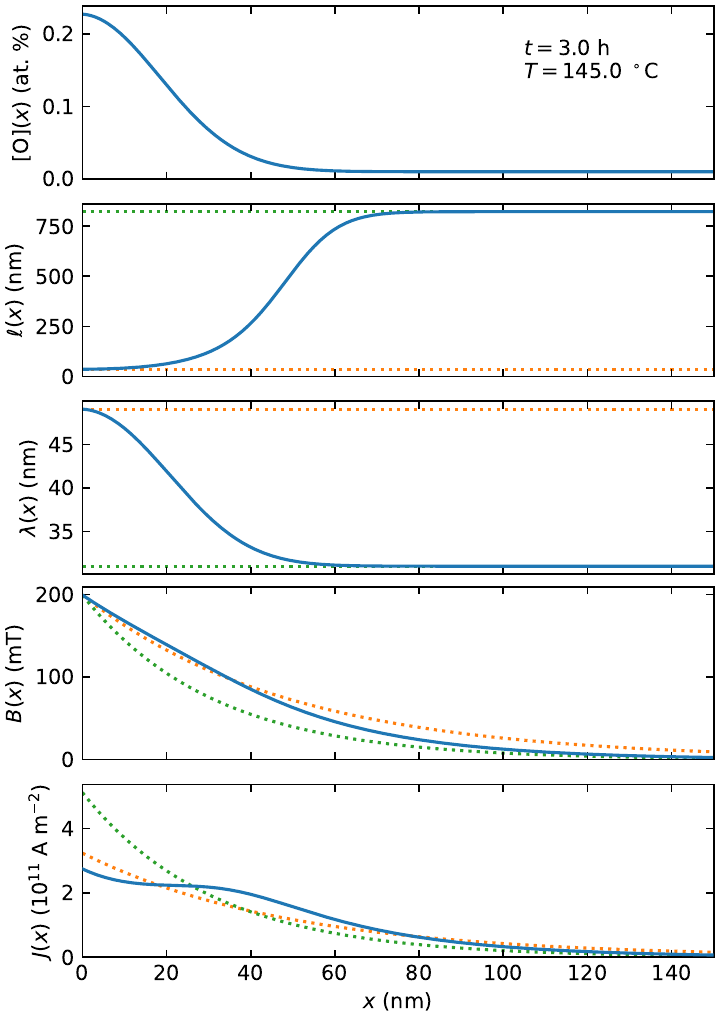} \hfill
    \includegraphics[width=\figtextfrac\textwidth]{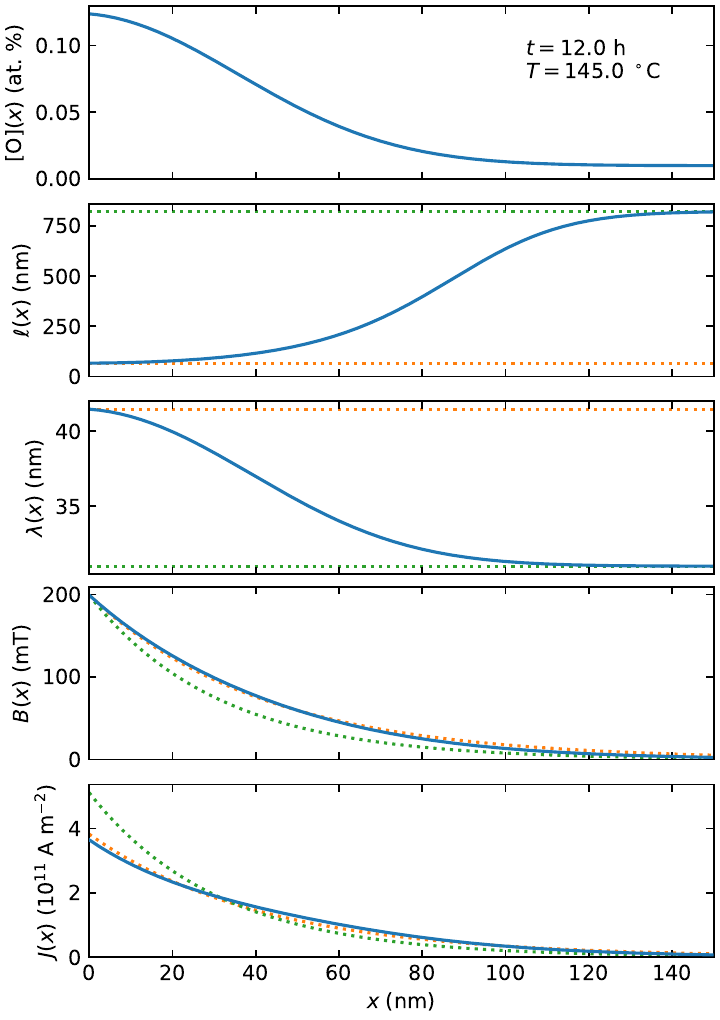} \hfill
    \includegraphics[width=\figtextfrac\textwidth]{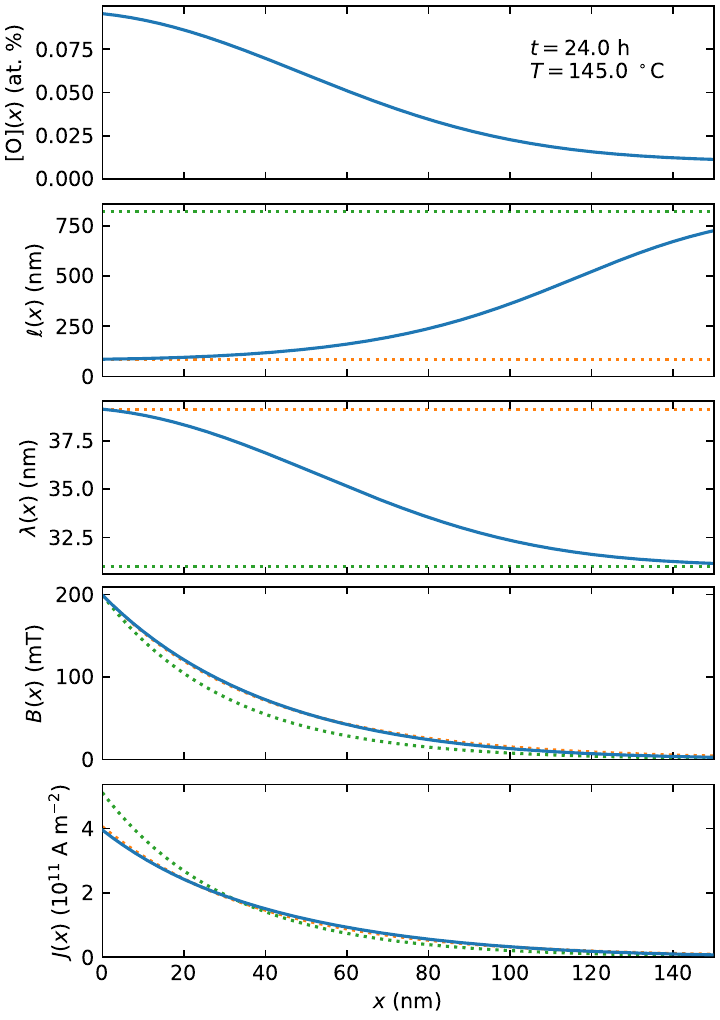} \hfill
    \includegraphics[width=\figtextfrac\textwidth]{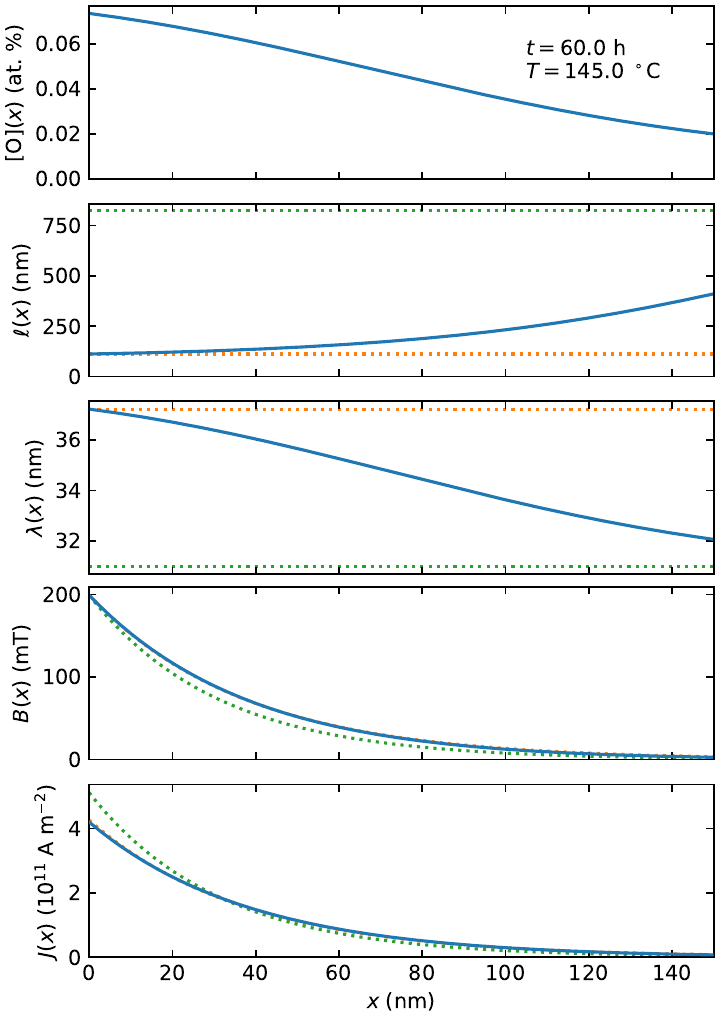}
    \caption{
        \label{fig:sim-145C}
        Simulated oxygen defect profile $[\ch{O}]$ as a function of depth $x$
        for several vacuum annealing treatment times $t$
        (indicated in each plot's top panel)
        at \qty{145}{\celsius}.
        The simulations and derived quantities
        make use of the parameters tabulated in \Cref{tab:sim-params}.
        The spatial form of the corresponding electron mean-free-path $\ell$
        [\Cref{eq:mfp-rrr,eq:residual-resistivity}],
        magnetic penetration depth $\lambda$
        [\Cref{eq:lambda-impurity,eq:lambda-two-fluid,eq:nlme}],
        Meissner screening profile $B(x)$
        [\Cref{eq:london-inhomogeneous}],
        and
        supercurrent density $J(x)$
        [\Cref{eq:supercurrent-density}],
        are also shown,
        revealing surface-localized inhomogeneities resulting from $[\ch{O}](x)$.
        For comparison,
        ``clean'' and ``dirty'' limits for the latter quantities,
        identified from $\ell$'s asymptotic limits,
        are shown as dotted green and orange lines,
        respectively.
    }
\end{figure*}

\begin{figure*}[hp]
    \centering
    \includegraphics[width=\figtextfrac\textwidth]{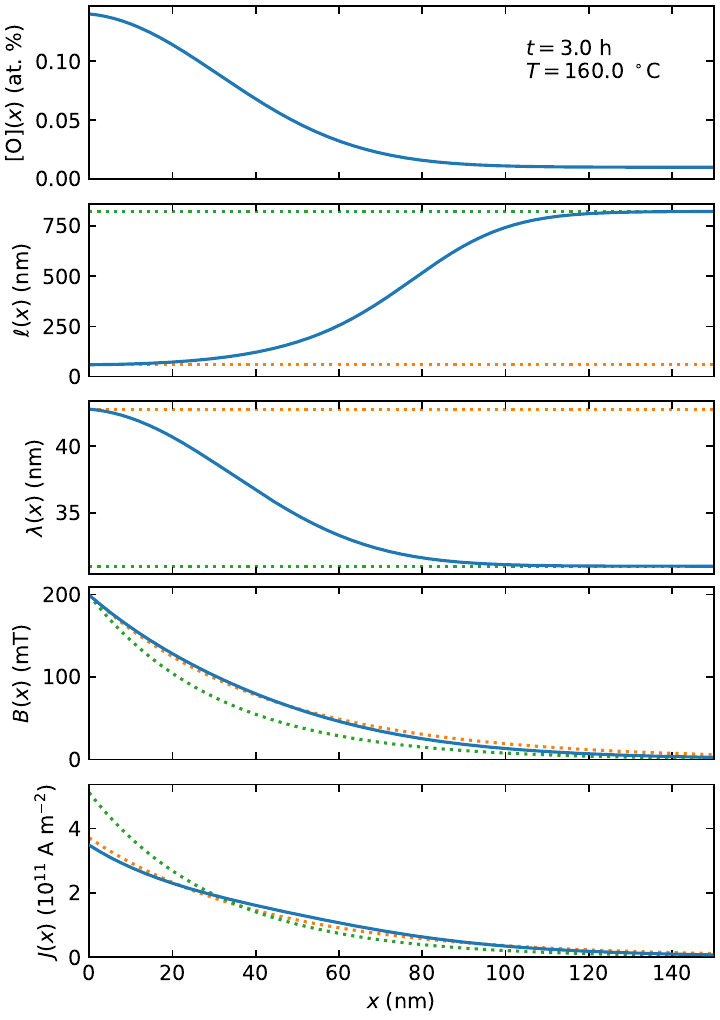} \hfill
    \includegraphics[width=\figtextfrac\textwidth]{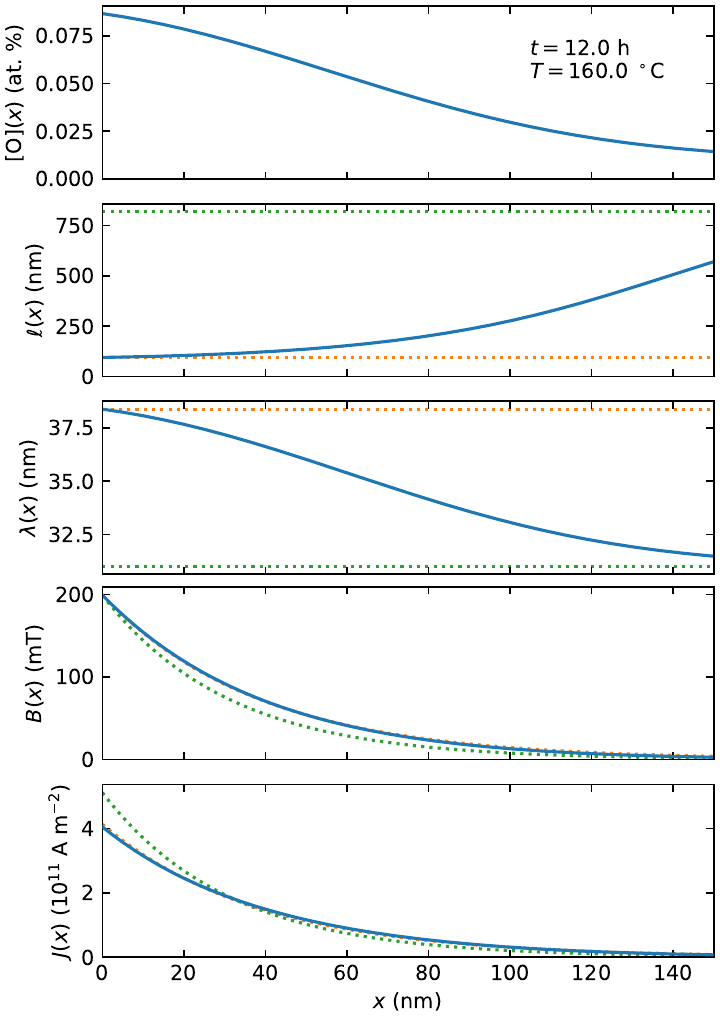} \hfill
    \includegraphics[width=\figtextfrac\textwidth]{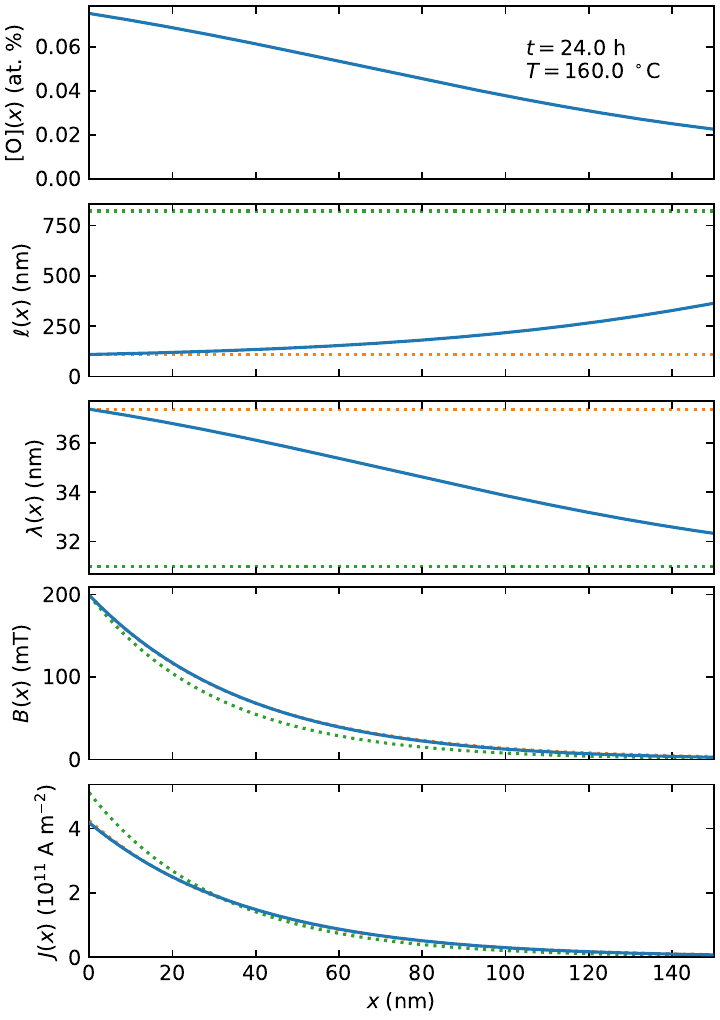} \hfill
    \includegraphics[width=\figtextfrac\textwidth]{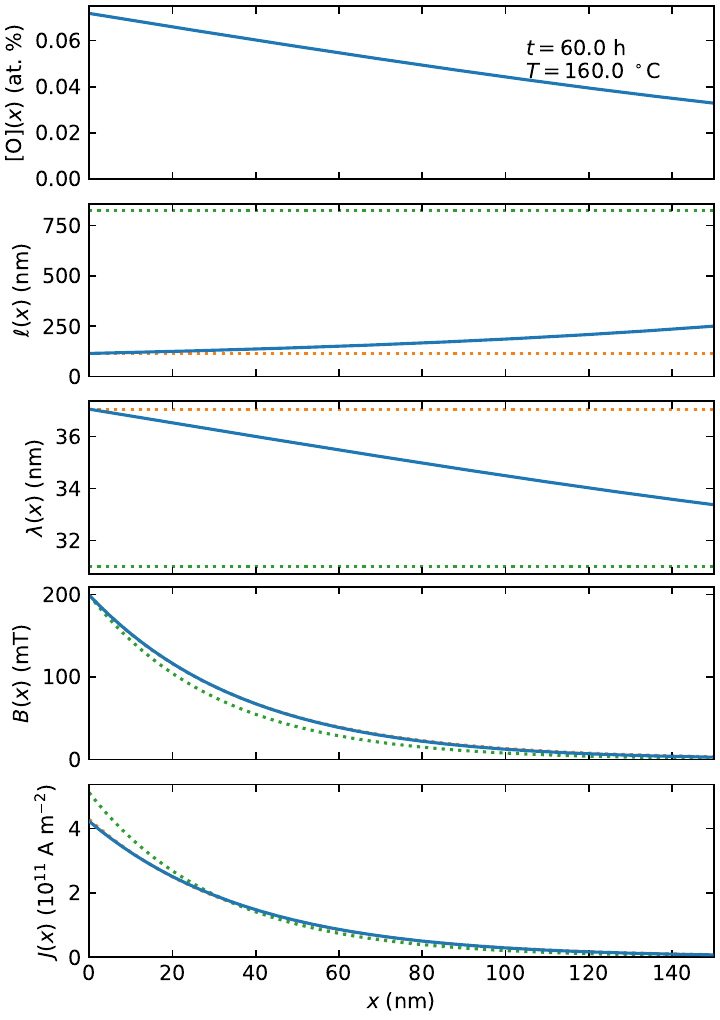}
    \caption{
        \label{fig:sim-160C}
        Simulated oxygen defect profile $[\ch{O}]$ as a function of depth $x$
        for several vacuum annealing treatment times $t$
        (indicated in each plot's top panel)
        at \qty{160}{\celsius}.
        The simulations and derived quantities
        make use of the parameters tabulated in \Cref{tab:sim-params}.
        The spatial form of the corresponding electron mean-free-path $\ell$
        [\Cref{eq:mfp-rrr,eq:residual-resistivity}],
        magnetic penetration depth $\lambda$
        [\Cref{eq:lambda-impurity,eq:lambda-two-fluid,eq:nlme}],
        Meissner screening profile $B(x)$
        [\Cref{eq:london-inhomogeneous}],
        and
        supercurrent density $J(x)$
        [\Cref{eq:supercurrent-density}],
        are also shown,
        revealing surface-localized inhomogeneities resulting from $[\ch{O}](x)$.
        For comparison,
        ``clean'' and ``dirty'' limits for the latter quantities,
        identified from $\ell$'s asymptotic limits,
        are shown as dotted green and orange lines,
        respectively.
    }
\end{figure*}

To complement the results described in \Cref{sec:results},
we provide additional simulation results at select baking times $t$
for common treatment temperatures $T$.
Specifically,
\Cref{fig:sim-110C} shows results at $T = \qty{110}{\celsius}$,
\Cref{fig:sim-120C} shows results at $T = \qty{120}{\celsius}$,
\Cref{fig:sim-145C} shows results at $T = \qty{145}{\celsius}$,
and
\Cref{fig:sim-160C} shows results at $T = \qty{160}{\celsius}$.
The $t$ and $T$ combinations chosen reflect common treatments ``recipes''
used for \gls{srf} cavities
(see, e.g.,~\cite{2006-Visentin-PC-441-66,2004-Ciovati-JAP-96-1591,2007-Visentin-SRF-13-304}).
From these examples,
there is a clear evolution to how the supercurrent density $J(x)$ is
modified spatially.
At $t$ and $T$ combinations that yield oxygen doping profiles
$[\ch{O}](x)$ localized close to the surface,
the strongest supercurrent deformations  are observed,
yielding peaks in the $J(x)$ profiles at shallow depths.
At constant $T$,
increasing $t$ has the effect of pushing the peak in $J(x)$ deeper
below the surface,
as well as decreasing its amplitude.
At sufficiently long $t$,
the peak vanishes and the supercurrent response approaches that of
(spatially homogeneous) ``dirty'' \ch{Nb}.
As $T$ increases,
the $t$ required for this to transpire decreases.
From this survey,
it is also evident that near equivalent doping effects are possible from very different
$t$ and $T$ combinations.
For example,
we point out the similarity of \qty{110}{\celsius}/\qty{60}{\hour} and
\qty{145}{\celsius}/\qty{3}{\hour} treatments,
consistent with empirical observations~\cite{2006-Visentin-PC-441-66}.

\bibliography{references.bib}

\end{document}